\documentstyle[12pt]{article}
\advance\textheight by \topskip
\begin{document}
\def\lsim{\mathrel{\lower2.5pt\vbox{\lineskip=0pt\baselineskip=0pt
\hbox{$<$}\hbox{$\sim$}}}}
\def\gsim{\mathrel{\lower2.5pt\vbox{\lineskip=0pt\baselineskip=0pt
\hbox{$>$}\hbox{$\sim$}}}}
\def\gs{SU(2)_{\rm L} \times U(1)_{\rm Y}}
\def\wh{{\cal W}}
\def\bh{{\cal B}}
\def\wtu{\wh^{\mu \nu}}
\def\wtd{\wh_{\mu \nu}}
\def\btu{\bh^{\mu \nu}}
\def\btd{\bh_{\mu \nu}}
\def\cl{{\cal L}}
\def\nll{\cl_{\rm NL}}
\def\ecl{\cl_{\rm EChL}}
\def\fpnl{\cl_{\rm FP}^{\rm NL}}
\def\fpl{\cl_{\rm FP}}
\def\msb{{\overline{\rm MS}}}
\def\mh{M_H}
\begin{titlepage}
\title{\Large{\bf Non--decoupling effects of the SM Higgs boson\\
to one loop}}
\author{Mar\'{\i}a J.
Herrero\thanks{e--mail:herrero@vm1.sdi.uam.es}\ \ \ and \
\ Ester Ruiz Morales\thanks{e--mail:
meruiz@vm1.sdi.uam.es} \\[3mm] Departamento de F\'{\i}sica
Te\'orica\\ Universidad Aut\'onoma de Madrid\\ Cantoblanco,\
\ 28049 -- Madrid,\ \ Spain} \date{}
\maketitle
\def\baselinestretch{1.3}
\begin{abstract}
\noindent
{\normalsize
We study the complete non-decoupling effects of the standard model
Higgs boson to one loop. Using effective field theory methods, we
integrate out the Higgs boson and represent its non-decoupling
effects by a set of gauge invariant effective operators of the
electroweak chiral Lagrangian.	In a previous work, we analyzed the
non-decoupling effects in the two and three-point Green's functions
of gauge fields. We complete here the calculation of the chiral
effective operators by analyzing the four-point functions.  We
discuss in detail the relation between the renormalization of both
the standard model and the effective theory, which is crucial for a
correct understanding and use of the electroweak chiral Lagrangian.
Some examples have been chosen to show the applicability of this
effective Lagrangian approach in the calculation of low energy
observables in electroweak theory.}
\end{abstract}

\vskip-18.5cm
\rightline{{\bf \large FTUAM 94/11}} \vspace{1mm}

\rightline{{\bf \large October 1994}}\vspace{1mm}

\rightline{{\bf \large hep-ph/9411207}}

\end{titlepage}

\newpage

\section{Introduction}

\renewcommand\baselinestretch{1.3}

Non-decoupling effects of heavy particles in the low energy
observables have been for a long time an indirect but crucial
test for the discovery of new particles. The most recent example
is the top quark, which contributes significantly to the
observables that measure the electroweak (EW) radiative
corrections, as for instance $\Delta \rho$ \cite{VELT1}. In
particular, the latest global fit of the electroweak parameters
from LEP data gives a quite strong constraint for the allowed
top mass value \cite{LEPTOP}. This indirect search supports the
recently announced evidence of $t {\bar t}$ production at CDF
\cite{CDFTOP} and will certainly contribute to get a final
confirmation of the existence of the quark top.

The non-decoupling effects of the Higgs boson are, however,
weaker than in the case of the top quark due to the screening
theorem \cite{VELT2}. According to this theorem, the sensitivity
in the low energy observables to the Higgs boson mass is at most
logarithmic at one loop. This fact has made in the past a
difficult task to disentangle the Higgs from the dominant
fermionic effects and, therefore, no significant bound on the
Higgs mass has been obtained so far from the analysis of LEP
data. This situation, however, will likely improve in the near
future due to the increasing precision of electroweak
measurements and the forthcoming CDF data.  A confirmation of
the top quark is still needed, as well as a precise measurement
of the top quark mass, but the important point is that it is now
beginning to be plausible the search for evidence of the Higgs
boson in electroweak radiative corrections.

On the other hand, regarding the theoretical aspects, the
leading logarithmic Higgs mass effects in the low energy
observables are known up to one-loop level since the pioneering
works by Appelquist and Bernard \cite{AB} and by Longhitano
\cite{LON}.  Their strategy was to use the symmetry properties
of the $\gs$ gauged non-linear $\sigma$-model (GNL) \cite{CHG}
along with a systematic power-counting analysis to provide a
list of these logarithmic Higgs mass dependent terms. However,
at present, it is known that these leading logarithmic terms are
not sufficient to discriminate a heavy Higgs possibility from an
alternative symmetry breaking scenario to which one requires to
respect the same SM symmetries. These logarithmic contributions
are a consequence of the general gauge and custodial symmetry
requirements of the low energy structure of EW interactions and
therefore they will be the same irrespective of the particular
choice for the breaking dynamics, with the Higgs mass being
replaced by some alternative physical mass.  Thus, if one wants
to reveal the nature of the symmetry breaking from low energy
observables, one has to go beyond the leading logarithmic
effects.

In this paper, we present the complete calculation of the
leading and next-to-leading non-decoupling effects of a heavy
Higgs boson in the SM to one loop level \footnote{For
simplicity, we will ignore the fermions in all the discussion
but their contribution, which is assumed here to be the SM one,
has to be added in any comparison with data.}.	These genuine
Higgs boson effects cannot be obtained using the GNL as in
\cite{AB,LON}, but must be calculated directly from the
evaluation of the one loop diagrams in the SM. Futhermore, in
order to classify in a systematic way those effects, we will use
here the electroweak chiral Lagrangian (EChL) \cite{HR2}. Our
approach is based on effective field theory methods, in which
the non-decoupling effects of a heavy Higgs boson are
represented, at energies below the Higgs mass, by certain set of
gauge invariant effective operators of the EChL.

The EChL is basically a non-linear sigma model coupled to the
$\gs$ gauge fields, where the Higgs field has been removed from
the physical spectrum of the theory. The model respects the
fundamental symmetries of the SM, namely $\gs$ gauge invariance
spontaneously broken to $U(1)_{\rm em}$ and the custodial
symmetry $SU(2)_{\rm C}$ of the pure scalar sector, but does not
include explicitely a particular dynamics for the symmetry
breaking \cite{HR2}-\cite{BDV}. Although this Higgs-less
parametrization of EW interactions is non-renormalizable, the
theory can be rendered finite to one loop by adding gauge
invariant operators up to dimension four. These effective
operators parametrize the low energy effects of the underlying
fundamental dynamics of the symmetry breaking.	In particular,
the EChL can be regarded as an effective theory of the SM in
which the Higgs field has been integrated out, and its effects
at energies well below the Higgs mass are parametrized by the
chiral effective operators \cite{HR2}.	We believe that this
kind of approach may be interesting for several reasons. First
of all, it provides a gauge invariant way of separating the
non-decoupling Higgs boson effects from the rest of the EW
radiative corrections.	On the other hand, the EChL is a general
framework in which one can analyze the low energy effects not
only of a heavy Higgs in the SM, but of more general breaking
dynamics characterized by the absence of ligth modes \cite{AW}.
It is then desirable to have the EChL that parametrizes a SM
Higgs as a fundamental reference model.

In our previous work \cite{HR2}, we discussed a general
procedure to obtain the EChL operators by matching the SM
predictions in the limit of large Higgs mass with the
predictions from the EChL to one loop order. The subset of EChL
operators involved in the two and three-point Green's functions
of the gauge fields were also obtained there. In this work, we
complete the calculation of the EChL operators by analyzing the
four-point Green's functions for gauge fields.  We will also
discuss here in more detail the relation between the
renormalization of both the SM and the effective theory, which
is crucial for understanding the true meaning of the effective
operators. We will show with some explicit examples how to
calculate observables in the EChL parametrization of EW
interactions.

The paper is organized as follows. Section two is a survey of
the electroweak chiral Lagrangian approach, in which we also
include a brief discussion of the formal renormalization
procedure of the effective theory and the matching conditions.
Section three is devoted to describe the computation of the SM
four-point Green's functions in the large $\mh$ limit.  The
renormalization prescription chosen for the SM is fixed in this
section to be the on-shell scheme. The matching equations will
be solved in section four where we present the results for the
complete set of EChL bare parameters. Section five is devoted to
discuss the dependence of the bare EChL parameters on the
renormalization prescription fixed for the underlying SM. We
explain in section six the relation between the bare EChL
parameters and the renormalized EChL parameters for different
renormalizations of the effective theory. In section seven we
explain how to compute observables to one loop with the EChL and
demonstrate by choosing some simple observables that the result
agrees with previous calculations in the literature. Finally,
section eight is devoted to the conclusions.

\section{The electroweak chiral Lagrangian}

The EChL is the most simple effective theory of EW interactions
that parametrizes the physics of the $\gs$ breaking dynamics at
low energies. The assumption made in this approach is that,
whatever the $\gs$ breaking interactions may be, the particles
involved in the symmetry breaking are heavier than the W and Z
bosons.  The EChL is then a low energy formulation of EW
interactions which contains just the "light" gauge and would-be
Goldstone fields, satisfying the basic requirement of $\gs$
gauge invariance spontaneously broken to $U(1)_{em}$:
\begin{equation}
\ecl = \nll + \sum_{i=0}^{13} \cl_{i}. \label{ECL}
\end{equation}
Its basic structure is a gauged non-linear sigma model $\nll$,
where a non-linear parametrization of the would-be Goldstone
bosons is coupled to the $\gs$ gauge fields
\begin{equation}
\nll  =  \frac{v^2}{4}\; Tr\left[ D_\mu U^\dagger D^\mu U \right]
+ \frac{1}{2}\; Tr\left[ \wtd \wtu + \btd \btu
\right] + \cl_{\rm R_\xi} + \fpnl,
\label{NLL}
\end{equation}
where the bosonic fields have been parametrized as
\begin{eqnarray}
U & \equiv & \displaystyle{\exp\left( {i \;
\frac{\vec{\tau}\cdot\vec{\pi}}{v}}\right)},\;\;\;
v  = 246 \;{\rm GeV}, \;\;\; \vec{\pi} = (\pi^1,\pi^2,\pi^3),
\nonumber\\
 \wh_\mu & \equiv & \frac{ -i}{2}\;
\vec{W}_\mu \cdot \vec{\tau}, \nonumber\\
\bh_\mu & \equiv & \frac{ -i}{2} \; B_\mu \;
\tau^3,\label{FPAR}
\end{eqnarray}
and the covariant derivative and the field strength tensors are
defined as
\begin{eqnarray}
D_\mu U & \equiv & \partial_\mu
U - g \wh_\mu U + g' U \bh_\mu, \nonumber\\[2mm]
\wtd & \equiv &  \partial_\mu \wh_\nu - \partial_\nu \wh_\mu -
g [\wh_\mu, \wh_\nu],\nonumber\\[2mm]
\btd & \equiv &  \partial_\mu \bh_\nu - \partial_\nu \bh_\mu.
\label{FTEN}
\end{eqnarray}
The physical fields are given by
\begin{eqnarray}
W^\pm_\mu & = & \frac{W^1_\mu \mp i W^2_\mu}{\sqrt 2},
\nonumber\\[2mm] Z_\mu & = & c\; W^3_\mu - s\; B_\mu
\nonumber, \\[2mm] A_\mu & = & s\; W^3_\mu + c\; B_\mu,
\end{eqnarray}
where $c = \cos \theta_{\rm w}, \; s = \sin
\theta_{\rm w}$ and the weak
angle is defined by $\tan \theta_{\rm w} = g' / g$.

The second term in eq.(\ref{ECL}) includes the complete set of
$\gs$ and CP invariant operators up to dimension four
\footnote{ There is an extra term $\cl_{14}$ proportional to
$\epsilon^{\mu\nu\alpha\beta}$ that is $CP$ conserving but $C$
and $P$ violating. It is not relevant in case of absence of
fermion contributions and will not be considered here.} that
were classified by Longhitano in \cite{LON}
\footnote{ The relation with Longhitano's notation is the
following: $
a_0=\frac{g^2}{g'^2}\beta_1\;;\;a_1=\frac{g}{g'}\alpha_1\;;\;
a_2=\frac{g}{g'}\alpha_2\;;\;a_3=-\alpha_3\;;\;a_i=\alpha_i\;,
i=4,5,6,7\;;\;a_8=-\alpha_8\;;\;a_9=-\alpha_9\;;\;
a_{10}=\alpha_{10}/2\;;\;a_{11}=\alpha_{11}\;;\;
a_{12}=\alpha_{12}/2\;;\;a_{13}=\alpha_{13}$.
Notice that the definition of $a_0$ is
different here than in \cite{HR2}.}:
\begin{eqnarray}
\cl_{0} & = & a_0 g'^2 \frac{v^2}{4} \left[ Tr\left(
T V_\mu \right) \right]^2 \nonumber\\[2mm]
\cl_{1} & = & a_1 \frac{i g g'}{2} B_{\mu\nu}
Tr\left( T \wtu \right) \nonumber\\[2mm]
\cl_{2} & = & a_2 \frac{i g'}{2} B_{\mu\nu}
Tr\left( T [V^\mu,V^\nu ] \right) \nonumber\\[2mm]
\cl_{3} & = & a_3  g Tr\left( \wtd [V^\mu,V^\nu ]\right)
\nonumber\\[2mm]
\cl_{4} & = & a_4  \left[ Tr\left( V_\mu V_\nu \right)
\right]^2 \nonumber\\[2mm]
\cl_{5} & = & a_5  \left[ Tr\left( V_\mu V^\mu \right)
\right]^2 \nonumber\\[2mm]
\cl_{6} & = & a_6 Tr\left( V_\mu V_\nu \right) Tr\left( T V^\mu
\right) Tr\left( T V^\nu \right)\nonumber\\[2mm]
\cl_{7} & = & a_7 Tr\left( V_\mu V^\mu \right) \left[
Tr\left( T V^\nu \right) \right]^2\nonumber\\[2mm]
\cl_{8} & = & a_8  \frac{g^2}{4} \left[ Tr\left( T \wtd \right)
\right]^2 \nonumber\\[2mm]
\cl_{9} & = & a_9  \frac{g}{2} Tr\left( T \wtd \right)
Tr\left( T [V^\mu,V^\nu ] \right) \nonumber\\[2mm]
\cl_{10} & = & a_{10} \left[ Tr\left( T V_\mu \right)
Tr\left( T V_\nu \right) \right]^2 \nonumber\\[2mm]
\cl_{11} & = & a_{11} Tr\left( ( D_\mu V^\mu )^2 \right)
\nonumber\\[2mm]
\cl_{12} & = & a_{12} Tr\left( T D_\mu D_\nu V^\nu \right)
Tr \left( T V^\mu \right)\nonumber\\[2mm]
\cl_{13} & = & a_{13} \frac{1}{2} \left[ Tr \left( T D_\mu V_\nu
\right) \right]^2 \label{Li}
\end{eqnarray}
where the basic building blocks are defined as
\begin{equation}
T  \equiv  U \tau^3 U^\dagger, \hspace{2cm}
 V_\mu\equiv(D_\mu U) U^\dagger .
\end{equation}
The above particular base of invariants can be transformed into
a new one with only 11 independent structures by making use of the
classical equations of motion \cite{MUL}. The new effective Lagrangian
is then given in terms of a new set of chiral parameters $\hat{a}_i$
given by: $\hat{a}_1 = a_1 + a_{13}$, $\hat{a}_4 = a_4 - a_{13}$,
$\hat{a}_5 = a_5 + a_{13}$, $\hat{a}_6 = a_6 - a_{13}$,
$\hat{a}_7 = a_7 + a_{13}$, $\hat{a}_8 = a_8 + a_{13}$,
$\hat{a}_{11}=\hat{a}_{12}=\hat{a}_{13}=0$; $ \hat{a}_i = a_i,
i=0,2,3,9,10$. Both effective Lagrangians, however, will give rise
to the same physical on-shell amplitudes. In this work, since
we will not restrict ourselves to calculate on-shell matrix elements,
we keep the complete basis given in eq.(\ref{Li}).

We will work in a generic R$_\xi$ gauge, the gauge fixing term
$\cl_{{\rm R}_\xi}$ and the Faddeev-Popov Lagrangian $\fpnl$ in
eq.(\ref{ECL}) were given in our previous work~\cite{HR2}. We refer
the reader to this work for the detailed formulas and a discussion on
these terms. It is worth just recalling here that $\fpnl$ does not
coincide with the usual Faddeev-Popov Lagrangian of the SM due to the
non-linearity of the would-be Goldstone bosons under infinitesimal
$\gs$ transformations. Furthermore, because of the non-linear
realization of the gauge symmetry, some of the couplings in
$\nll$ have different Feynman rules than in the SM. We collect
in fig.(1) the subset of them that are relevant for the present
calculation.

It is also important to mention that in a R$_\xi$ gauge, the
complete electroweak chiral Lagrangian must be BRS invariant and
include also effective operators involving the ghost fields.
Longhitano\cite{LON} showed, however, that the subset of
operators given above is sufficient to absorb the divergences of
the gauged non-linear sigma model if the Landau gauge is chosen.
As it will be shown later, we have demonstrated in this work  that
the set of operators given in eq.(\ref{Li}) is also enough, in a
general  R$_\xi$ gauge, to render finite the two, three and
four-point Green's functions with external gauge fields.

The non-linear sigma model in eq.(\ref{NLL}) is not a
renormalizable theory, as increasing the number of loops in a
calculation implies the appearence of new divergent structures
of higher and higher dimension. However, the EChL is an
effective theory that can be renormalized order by order in the
loop expansion.  In particular, at one loop order, the new
divergences generated by a one loop calculation with $\nll$  can
be absorved into redefinitions of the effective operators given
in eq.(\ref{Li}) \cite{LON}.  Therefore, one can obtain finite
renormalized Green's functions if one makes a suitable
redefinition of the fields and parameters of the EChL, among
which the chiral parameters $a_i$ must be included \cite{W}-
\cite{GL2}. Formally, one defines the renormalized quantities in the
effective theory by the following relations
\begin{center}$
\begin{array}{ll}
B_{\mu}^b  \; = \; \widehat{Z}_B^{1/2} \; B_\mu, \hspace{1cm}& g'^b
\; =\; \widehat{Z}_B^{-1/2}\; ( g' - \widehat{\delta g'} ), \\[2mm]
\vec{W}_\mu^b \; =\;  \widehat{Z}_W^{1/2}\; \vec{W}_\mu, &
 g^b \; =\; \widehat{Z}_W^{-1/2}\; ( g - \widehat{\delta g} ),
\\[2mm]
\vec{\pi}^b \; =\; \widehat{Z}_\Phi^{1/2}\; \vec{\pi}, &
v^b \; = \; \widehat{Z}_\Phi^{1/2}\; ( v -\widehat{ \delta v}),\\[2mm]
\xi_B^b \; = \; \xi_B\; ( 1 + \widehat{\delta \xi}_B), \hspace{1cm}&
\xi_W^b \; = \; \xi_W \;( 1 + \widehat{\delta \xi}_W),
\end{array}$
\begin{equation}
a_i^b \; = \; a_i(\mu) \; + \; \delta a_i , \label{RET}
\end{equation}
\end{center}
where the renormalization constants of the effective theory are
$ \widehat{Z}_i \equiv 1 + \widehat{\delta Z_i}$ and the
superscript b denotes bare quantities. We use the hatted
notation to distinguish counterterms and Green's functions in the
effective theory from the corresponding quantities in the SM.

The 1PI renormalized Green's functions of the effective theory
to one loop will be generically denoted by
\begin{equation}
\widehat{\Gamma}^{\rm R} = \widehat{\Gamma}^{\rm T} +
\widehat{\Gamma}^{\rm C} + \widehat{\Gamma}^{\rm L},
\label{GFE}
\end{equation}
where the superscript R denote renormalized function and
the superscripts T, C and L denote the tree level, counterterm
and loop contributions respectively.  We will
discuss in section 6 the on-shell renormalization of the effective
theory, giving explicit expressions for the counterterms
introduced in eq.(\ref{RET}). For the moment, in order to discuss the
matching procedure, we will treat the counterterm contributions to
the renormalized functions of eq.(\ref{GFE}) just at a formal level.

We would like to focus now our attention on the chiral
parameters.  Once a particular renormalization scheme has been
chosen to fix the counterterms of the effective theory, the
renormalized $a_i(\mu)$ parameters remain as free parameters
that can not be determined within the framework of the low
energy effective theory.  The values of the renormalized chiral
parameters can be constrained from the experiment, as they are
directly related to different observables in scattering
processes \cite{DH,BDV},\cite{DHT}-\cite{MAY} and in precision
electroweak measurements (\cite{HT}-\cite{EH},\cite{PES}, see also section
7); but to have any theoretical insight on their values, one has
to relate the effective theory with a particular underlying
fundamental dynamics of the symmetry breaking.

If the underlying fundamental theory is the standard model with a
heavy Higgs, the chiral parameters can be determined by matching the
predictions of the SM in the large Higgs mass limit and those of the
EChL, at one loop level.  By heavy Higgs we mean a Higgs mass much
larger than any external momenta and light particle masses ($ p^2,
M_Z^2 \ll \mh^2$) so that one can make a low energy expansion, but
smaller than say 1 TeV, so that a perturbative loop calculation is
reliable.

We will impose here the strongest form of matching \cite{GEO} by
requiring that all renormalized one-light-particle irreducible (1LPI)
Green's functions are the same in both theories at scales $\mu \leq
\mh$. The 1LPI functions are, by definition, the Green's functions
with only light particles in the external legs and whose contributing
graphs cannot be disconnected by cutting a single light particle
line.  This matching condition is equivalent to the equality of the
light particle effective action in the two descriptions.
Some other forms of matching have been discussed in the literature,
by requiring the equality of the two theories at the level of
the physical scattering amplitudes \cite{DH} or connected Green's
functions \cite{DOM}. These requirements, however, complicate the
calculation unnecesarily while give at the end the same results
for the physical observables.
There is also some discussion in the literature \cite{DOM} on the
dependence of the Green's functions on the parametrization
chosen for the would-be Goldstone bosons. We will fix here the particular
parametrizations of eq.(\ref{FPAR}) in the effective theory
and eq.(\ref{SMPAR}) in the SM. Of course, the physical observables
will not depend on this particular choice.

In order to completely determine the chiral parameters in terms of
the parameters of the SM, it is enough to impose matching conditions
in the two, three, and four-point 1LPI renormalized Green's functions
with external gauge fields.  We have worked in a general
R$_\xi$-gauge to show that the chiral parameters $a_i$ are
$\xi$-independent. We use dimensional regularization and the on-shell
substraction scheme.

The SM Green's functions are non-local; in particular, they depend on
$p / \mh$ through the virtual Higgs propagators. One has to make a
large $M_H$ expansion to represent the virtual Higgs boson effects by
the local effective operators ${\cal L}_i$. In this step, care must
be taken since clearly the operations of making loop integrals and
taking the large $\mh$ limit do not commute. Thus, one must first
regulate the loop integrals by dimensional regularization, then
perform the renormalization with some fixed prescription (on-shell in
our case) and at the end take the large $\mh$ limit, with $\mh$ being
the renormalized Higgs mass.  From the computational point of view,
in the large $\mh$ limit we have neglected  contributions that depend
on $(p/\mh)$ and/or $(M_V/\mh, M_V = M_W, M_Z))$ and vanish when the
formal $\mh \rightarrow \infty$ limit is taken. We show in appendix A
one illustrative example of how to take the large $\mh$ expansion of
the loop integrals.

The matching procedure can be summarized by the following relation
among renormalized 1LPI Green's functions
\begin{equation}
\Gamma^{\rm R}_{\rm SM} (\mu ) \; = \;
\widehat{\Gamma}^{\rm R}_{\rm EChL}
(\mu ) \; , \;\;\;\;\;\;\;\;\;\mu \leq \mh,
\label{match}
\end{equation}
where the large Higgs mass expansion of the SM Green's functions has
to be made as explained above. This matching condition imposes a
relation between the renormalization of the SM and the
renormalization of the effective theory. We have chosen to
renormalize both theories in the on-shell scheme, so that the
renormalized parameters are the physical masses and coupling
constants. Therefore, the renormalized parameters are taken to be the
same in both theories and the matching conditions will provide
relations between the SM and the EChL counterterms
\footnote{ In some related literature on effective field theories
\cite{GEO,SANTA}, the choice of a mass-independent substraction
prescription ($\msb$) in both theories has also been discussed. In
that case, the matching procedure relates the running
$\msb$-renormalized parameters, that are different in the fundamental
and the effective theories.}.

The matching condition (\ref{match}) represents
symbolically a system of tensorial coupled equations (as many as 1LPI
functions for external gauge fields) with several unknowns, namely
the complete set of parameters $a_i^b$ that we
are interested in determining. In our previous work \cite{HR2}, we
solved the subset of coupled equations involving the two-point and
three-point functions. From this subset, we were able to determine
the chiral parameters $a_0^b, a_1^b, a_2^b, a_3^b, a_8^b, a_9^b,
a_{11}^b, a_{12}^b$ and $a_{13}^b$. In section 4, we will solve
the matching equations for the four-point Green's functions, thus
the set of EChL parameters for a heavy Higgs will be completed.
But before that, we have to set a renormalization prescription
for the standard model.

\section{Renormalization of the standard model}

We start by writing down the SM Lagrangian
\begin{equation}
\cl_{\rm SM} = (D_\mu \Phi)^\dagger (D^\mu \Phi) + \mu^2
\Phi^\dagger \Phi - \lambda (\Phi^\dagger \Phi)^2 +
\frac{1}{2} Tr \left( \wtd \wtu + \btd \btu \right) +
\cl_{{\rm R}_\xi} + \cl_{\rm FP} ,\\[2mm]
\end{equation}
where
\begin{eqnarray}
\Phi & = & \frac{1}{\sqrt 2}\left(
\begin{array}{c}\phi_1 - i \phi_2 \\
 \sigma + i \chi \end{array}\right), \hspace{2cm}
(\pi_1,\pi_2,\pi_3)  \equiv  (-\phi_2,\phi_1,-\chi) ,
\nonumber\\[2mm]
D_\mu \Phi &  = &  ( \partial_\mu + \frac{1}{2} i g
\vec{W}_\mu\cdot\vec{\tau} +
\frac{1}{2} i g' B_\mu) \Phi . \label{SMPAR}
\end{eqnarray}
$\wtd, \btd$ are defined in eqs.(\ref{FPAR},\ref{FTEN}),
$\cl_{{\rm R}_\xi}$ and $\cl_{\rm FP}$ are the usual R$_{\xi}$
gauge fixing and Faddeev--Popov terms of the standard model.

We rescale the fields and parameters as follows
\begin{equation}
\begin{array}{ll}
B_{\mu}^b  = Z_B^{1/2} B_\mu , \hspace{2cm} &
\vec{W}_\mu^b  =  Z_W^{1/2} \vec{W}_\mu ,\\[2mm]
\Phi^b = Z_\Phi^{1/2} \Phi ,&
v^b  = Z_\Phi^{1/2} ( v - \delta v ) , \\[2mm]
g^b  = Z_W^{-1/2} ( g - \delta g ) , &
g'^b  =  Z_B^{-1/2} ( g' - \delta g' ) ,\\[2mm]
\mu^b  =  Z_\Phi^{-1/2} ( \mu - \delta \mu ) ,&
\lambda^b  =  \lambda (1 - \delta \lambda / \lambda) ,\\[2mm]
\xi_B^b  =  \xi_B ( 1 + \delta \xi_B ) , &
\xi_W^b  =  \xi_W ( 1 + \delta \xi_W ) .
\end{array} \label{REPS}
\end{equation}
where the renormalization constants of the SM are
$ Z_i \equiv 1 + \delta Z_i$ and the superscript b denotes bare
quantities.

We have chosen to renormalize the SM in the
on-shell scheme. We choose the physical masses, $\mh$, $M_W$,
$M_Z$ and $g$ as our renormalized parameters.
The weak mixing angle is defined in terms of physical
quantities, as it is usual in the on-shell scheme
\begin{equation}
\cos^2 \theta_W \equiv \frac{M_W^2}{M_Z^2}
\end{equation}
and from $g$ and $\theta_W$ one derives the coupling constant
$g' = g \tan \theta_W$.

The 1LPI renormalized Green's functions in the standard model
to one loop will be generically denoted by
\begin{equation}
\Gamma^{\rm R} = \Gamma^{\rm T} + \Gamma^{\rm C} + \Gamma^{\rm L},
\end{equation}
where one has to consider the tree, counterterm and loop
contributions of all the one light particle irreducible diagrams in
the SM; that is, all the diagrams that cannot be disconnected by
cutting a light (non-Higgs) particle line.

In principle, we should give now the whole set of renormalization
conditions defining the SM on-shell scheme. However, to extract from
the matching conditions the values of the chiral parameters $a_i$, we
only need for the moment to evaluate explicitely the SM counterterms
that enter in the renormalization of the diagrams $T_i$ in figs.(3),
that is, the tree level diagrams with an intermediate Higgs boson.
Furthermore, since we are doing a large $\mh$ expansion, it will not
be necessary to give the complete expressions for these SM
counterterms, but just the leading terms that give
non-negligible contributions (i.e. non vanishing in the large $\mh$
limit) once they are plugged into the matching equations of the
four-point functions. In summary, these considerations imply that we will
need explicit expressions for the tadpole and Higgs mass counterterms
to order  $\mh^4$ and for the  W and Z mass counterterms to order
$\mh^2$. The other SM counterterms, $\delta Z_W$ and $\delta g$,
have at most a logarithmic dependence on Higgs mass and give
subleading contributions to the renormalization of the $T_i$
diagrams. The $\delta Z_W$ and $\delta g$ counterterms, however,
do contribute to the renormalization of the four-point Green's
functions through the renormalization of the tree level irreducible
diagrams. We do not need to give explicit expressions for them
because, as we will see, they appear in the matching through the
differences $\Delta Z_W = \delta Z_W - \widehat{\delta Z}_W$ and
$\Delta g = \delta g - \widehat{\delta g}$. The values of these
differences will be extracted from the matching.

The renormalization of the scalar sector has been done following the
work of Marciano and Willenbrock~\cite{MAW}.
In order to fix the notation and renormalization for the tadpole
we first write down the SM Lagrangian for the scalar sector in
terms of the would-be Goldstone boson fields $\phi^\pm \equiv
(\phi_1 \mp i \phi_2) / \sqrt{2}$ and $\chi$ and the physical
Higgs boson field $H$. In terms of the bare fields and
parameters, it reads as follows
\begin{eqnarray}
\cl_{\rm SM}^{\rm scalar} & = &
\partial_\mu \phi^{+ b} \partial^\mu \phi^{- b} + \frac{1}{2}
\partial_\mu \chi^b \partial^\mu \chi^b + \frac{1}{2}
\partial_\mu H^b \partial^\mu H^b \nonumber\\[2mm]
& & - \lambda^b \left[ \phi^{+ b} \phi^{- b} (\phi^{+ b} \phi^{- b} +
(\chi^b)^2 + (H^b)^2 ) + \frac{1}{4} ( (\chi^b)^2 + (H^b)^2 )^2
\right] \nonumber\\
& & - \lambda^b v^b \left[ 2 \phi^{+ b} \phi^{- b} H^b +  (\chi^b)^2
H^b + (H^b)^3 \right] \nonumber \\[2mm]
& & - \lambda^b (v^b)^2 (H^b)^2 + \frac{\delta T}{v^b}
(\phi^{+ b} \phi^{- b} + \frac{1}{2} (\chi^b)^2 + \frac{1}{2} (H^b)^2 )
+ \delta T H^b \nonumber\\[2mm]
& & + ( \xi {\rm -dependent \;\; terms } )
\label{LESC}
\end{eqnarray}
The tadpole counterterm is defined in terms of bare
parameters as \footnote{ Notice that the sign chosen in the
definition of $\delta T$ is opposite to the one in \cite{MAW}}
\begin{equation}
\delta T \equiv v^b \left( (\mu^b)^2 - \lambda^b (v^b)^2 \right)
\end{equation}
The renormalization condition for the tadpole is fixed such that
the tadpole loop corrections $T$ are cancelled by the tadpole
counterterm $\delta T$, or equivalently, such that the
renormalized tadpole vanishes. From the computational point of
view, one ignores all tadpole diagrams and tadpole counterterms,
and includes the extra contributions to the renormalized scalar
Green's functions coming from the counterterm $\delta T / v$
in eq.(\ref{LESC}) that is quadratic in the scalar fields.

The bare masses of the Higgs and gauge bosons are taken as
\begin{equation}
(M_H^b)^2  =  2 \; \lambda^b \; (v^b)^2, \hspace{8mm}
(M_W^b)^2  =  (g^b)^2 \; (v^b)^2 / 4, \hspace{8mm}
(M_Z^b)^2  =  \left( (g^b)^2 + (g^{\prime b})^2 \right)\; (v^b)^2 / 4
\label{MWH}
\end{equation}
so that the following relation among the basic bare parameters holds
\begin{equation}
\lambda^b = \frac{(g^b)^2 \; (M_H^b)^2}{8 \; (M_W^b)^2} \label{LAMB}
\end{equation}
The renormalized masses are defined by
\begin{equation}
(M_H^b)^2  =  M_H^2 + \delta M_H^2, \hspace{8mm}
(M_W^b)^2  =  M_W^2 + \delta M_W^2, \hspace{8mm}
(M_Z^b)^2  =  M_Z^2 + \delta M_Z^2,
\end{equation}
so that the on-shell renormalization conditions
\begin{eqnarray}
& {\rm Re} \left[ \Sigma_{H}^{\rm R} (q^2 = \mh^2) \right]
 = 0,\hspace{1cm} &  T^{\rm R} \equiv T + \delta T = 0,\nonumber\\[2mm]
& {\rm Re} \left[ \Sigma_{W}^{\rm R} (q^2 = M_W^2) \right]
 = 0, \hspace{1cm}
& {\rm Re} \left[ \Sigma_{Z}^{\rm R} (q^2 = M_Z^2) \right]
= 0, \label{RCON}
\end{eqnarray}
imply that eqs.(\ref{MWH},\ref{LAMB}) are also fulfilled by the renormalized
quantities
\begin{eqnarray}
M_H^2 & = & 2  \lambda v^2, \nonumber\\[2mm]
M_W^2 & = & g^2 v^2 / 4, \nonumber\\[2mm]
M_Z^2 & = & (g^2 + g'^2) v^2 / 4, \nonumber\\[2mm]
\lambda & = & \frac{ g^2 M_H^2}{8 M_W^2}. \label{CORE}
\end{eqnarray}
The renormalization conditions (\ref{RCON}) fix the
values of the SM counterterms to be\footnote{ We
denote by $- i g_{\mu \nu} \Sigma_{\rm V}$, (V=W,Z)  and
$- i \Sigma_{\rm H}$ the direct result from the
Feynman diagrams. The tadpole
loops are however denoted by $i T$.}
\begin{equation}
\begin{array}{ll}
 \delta \mh^2  =  - {\rm Re} \left[ \Sigma_{H}^{\rm L}
(q^2 = \mh^2) \right] +{\displaystyle \frac{\delta T}{v}}, \hspace{1cm}
& \delta T  =  - T, \\[4mm]
 \delta M_W^2  =  {\rm Re} \left[ \Sigma_{W}^{\rm L}
(q^2 = M_W^2) \right], \hspace{1cm}
& \delta M_Z^2 =  {\rm Re} \left[ \Sigma_{Z}^{\rm L}
(q^2 = M_Z^2) \right]. \vspace{2mm}
\end{array} \label{RC2}
\end{equation}
If one wishes to keep just the non-vanishing
contributions in the large $\mh$ limit to the renormalization
of the tree level diagrams $T_i$, the computation of the unrenormalized
self-energies of eq.(\ref{RC2}) involve just the leading
diagrams collected in fig.(2). These loop diagrams give
the following values of the mass and tadpole counterterms in the
on-shell scheme
\begin{eqnarray}
\frac{\delta M_H^2}{M_H^2} & = & \frac{g^2 M_H^2}{M_W^2}
 \frac{  1}{16 \pi^2} \left[ \frac{3}{2} \hat{\Delta}_\epsilon
+ 3 - \frac{3 \sqrt{3}}{8} \pi \right] + {\cal O}(1),\nonumber\\[2mm]
\frac{\delta M_W^2}{M_W^2} & = & \frac{g^2 M_H^2}{M_W^2}
\frac{- 1}{16 \pi^2} \frac{1}{8} + {\cal O}(1),\nonumber\\[2mm]
 \frac{\delta M_Z^2}{M_Z^2} & = & \frac{g^2 M_H^2}{M_W^2}
\frac{- 1}{16 \pi^2} \frac{1}{8}  + {\cal O}(1),
\nonumber\\[2mm]
\frac{\delta T / v}{\mh^2} & = &  - \frac{g^2 M_H^2}{M_W^2}
\frac{ 1}{16 \pi^2} \frac{3}{8} \left[ \hat{\Delta}_\epsilon
+ 1 \right] + {\cal O}(1), \label{SMCL}
\end{eqnarray}
where
\begin{equation}
\hat{\Delta}_\epsilon  =  \Delta_\epsilon -
\log \frac{\mh^2}{\mu^2}, \hspace{1cm}
\Delta_\epsilon  =  \frac{2}{\epsilon} - \gamma_{\rm E} +
\log 4 \pi, \hspace{1cm} \epsilon = 4 - D
\label{EPS}
\end{equation}
and $\mu$ is the usual mass scale of dimensional regularization.
We have checked the agreement of our expressions
with the results of \cite{MAW}.

We have explicitely indicated in the formulas of the SM counterterms
(\ref{SMCL}) that these expressions are truncated to a certain
order in the 1/$\mh$ expansion. This truncation is enough to
keep the non-vanishing effects of a heavy Higgs in the evaluation
of the following combination of SM counterterms
\begin{equation}
\delta S = \frac{M_W^2}{g^2 M_H^2} \left(- \frac{\delta M_H^2}{M_H^2}
+ \frac{ \delta T / v}{M_H^2} + \frac{\delta M_W^2}{M_W^2} \right)
\label{DELS}
\end{equation}
that comes from the renormalization of the tree level diagrams $T_i$
and appears explicitely in the matching equations for the
four-point functions given in appendix B\footnote{In the two-point
functions however, it is necessary to go to the next order in the
large $\mh$ expansion of these terms \cite{HR2}.}.

\section{Matching equations for the 4-point Green's functions}

In this section we present the results of our calculation of the
4-point Green's functions, giving the set of matching
equations that we have imposed and their solution.
The master equation that summarizes the complete
set of matching conditions for the renormalized four-point functions is
the following:
\begin{equation}
 M_{abcd}^{{\rm T} \; \mu \nu \rho \lambda} +
 M_{abcd}^{{\rm C} \; \mu \nu \rho \lambda} +
 M_{abcd}^{{\rm L} \; \mu \nu \rho \lambda} =
 \widehat{M}_{abcd}^{{\rm T} \; \mu \nu \rho \lambda} +
 \widehat{M}_{abcd}^{{\rm C} \; \mu \nu \rho \lambda} +
 \widehat{M}_{abcd}^{{\rm L} \; \mu \nu \rho \lambda},
\label{MAMA}
\end{equation}
where $abcd = \gamma\gamma WW, \gamma ZWW, ZZWW, WWWW, ZZZZ$.

The calculation of the one loop contributions
$M_{\rm L}$ and $\widehat{M}_{\rm L}$ is the most involved part.
One must include all the 1PI diagrams
in the EChL and all the 1LPI diagrams in the SM.
1LPI diagrams are those that cannot be disconnected by cutting a
single light particle line, that is, a non-Higgs particle line.
One must, in principle, account for all kind of diagrams with
gauge, scalar and ghost fields flowing in the loops.
However, some simplifications occur.
Firstly, a subset of the diagrams that have only light particles in
it is exactly the same in both models, and their contribution can be
simply dropped out from both sides of the matching equation
(\ref{MAMA}). This is the case, for instance, of the subset of
diagrams whith only gauge particles in them.  Secondly, calculating
explicitely every diagram in the four-point 1LPI SM functions and using
the techniques given in app.A, we have checked that the diagrams
involving both gauge and Higgs particles in the loops give
vanishing contributions in the large $\mh$ limit to the four-point
functions. Only those with just scalars (Goldstone bosons or Higgs)
particles in the loops do contribute with non-vanishing corrections
in the large $\mh$ limit to the matching equation
(\ref{MAMA})\footnote{This is also true for the three-point functions but
it is false for the two-point functions where both pure scalar and
mixed gauge-scalar loops contribute in the large $M_H$ limit.}.
Finally, among the diagrams with pure scalar loops, there are some
with only Goldstone boson particles. One would expect that these
diagrams give the same contributions in the SM and the EChL, however
they do not.  The reason is the already mentioned differences in the
Feynman rules of the vertices in fig.(1). These diagrams (denoted
generically by $D_i$ in figs.(3) ) must therefore be included in both
sides of the matching equation.

In fig.(3), we give the complete list of the tree level ($T_i$) and
one loop ($L_i, D_i$) diagrams that give a
contribution to the matching equations (\ref{MAMA}) of the
$\gamma\gamma WW, \gamma ZWW, ZZWW,WWWW$ and $ZZZZ$ Green's functions.

We give in appendix B the final result of the calculation of the
tree, counterterms and loop contributions to the matching equations
for the different Green's functions.  Each matching condition of the
form given in eq.(B1) is a tensorial equation and therefore it
provides several equations corresponding to the various independent
tensorial structures. Furthermore, each of these equations can be
written in the form of a polynomial in powers of $c^2$. In summary,
one gets one equation "per" coefficient of the polinomial in each
independent tensorial structure and in each Green's function.
The result including the equations from the two, three and four-point
functions is a linear system with  more equations than unknowns.
The system turns out to be compatible, giving a strong consistency
check of the calculation. By keeping just the independent
set of equations from the four ponit functions, one gets:
\begin{flushleft} \vspace{-1cm}
\begin{eqnarray}
\left( \Delta Z_W - \frac{\Delta g^2}{g^2} \right) \frac{1}{g^2} &=&
\frac{1}{16 \pi^2} \frac{- 1}{12} \left[ \hat{\Delta}_\epsilon
+ \frac{5}{6} \right]\\[2mm]
\left( \Delta Z_W - \frac{\Delta g^2}{g^2} \right) \frac{1}{g^2}
+ a^b_{11} &=&
\frac{1}{16 \pi^2} \frac{- 1}{12} \left[ \hat{\Delta}_\epsilon
+ \frac{4}{3} \right] \nonumber \\[2mm]
2 a^b_3 &=& \frac{1}{16 \pi^2} \frac{- 1}{12} \left[
\hat{\Delta}_\epsilon + \frac{17}{6} \right] \nonumber \\[2mm]
a^b_3 - a^b_{11} + a^b_{12} &=&
\frac{1}{16 \pi^2} \frac{- 1}{24} \left[ \hat{\Delta}_\epsilon
+ \frac{11}{6} \right] \nonumber \\[2mm]
2 \left( a^b_5 + a^b_7 \right) &=&
\frac{1}{16 \pi^2} \frac{1}{12} \left[ \frac{43}{2}
\hat{\Delta}_\epsilon + \frac{47}{3} \right]
+ \frac{M_W^2}{g^2 M_H^2} + \delta S \nonumber \\[2mm]
a_4^b + a^b_6 - a^b_{11} + 2 a^b_{12} &=&
\frac{1}{16 \pi^2} \frac{- 1}{12} \left[ \hat{\Delta}_\epsilon
+ \frac{7}{3} \right] \nonumber \\[2mm]
\left( \Delta Z_W - \frac{\Delta g^2}{g^2} \right) \frac{1}{g^2}
 + 2 a^b_3 - a_4^b - a^b_8 + 2 a^b_9 - 2 a^b_{13} &=&
\frac{1}{16 \pi^2} \frac{- 1}{12} \left[
\hat{\Delta}_\epsilon + \frac{5}{6} \right] \nonumber \\[2mm]
\left( \Delta Z_W - \frac{\Delta g^2}{g^2} \right) \frac{1}{g^2}
 + 2 a^b_3 + a_4^b + 2 a^b_5 - a^b_8 + 2 a^b_9 - 2 a^b_{13} &=&
\frac{1}{16 \pi^2} \frac{1}{24} \left[
37 \hat{\Delta}_\epsilon + \frac{55}{3} \right]
+ \frac{M_W^2}{g^2 M_H^2} + \delta S \nonumber \\[2mm]
2 \left( a^b_4 + a^b_5 + 2 ( a^b_6 + a^b_7 + 2 a^b_{10})
\right) &=& \frac{1}{16 \pi^2} \frac{ 1}{8} \left[
13 \hat{\Delta}_\epsilon + \frac{20}{3} \right]
+ \frac{M_W^2}{g^2 M_H^2} + \delta S \nonumber
\end{eqnarray}
\end{flushleft}
where $\hat{\Delta}_\epsilon$ and $\delta S$ have been defined in eqs.
(\ref{EPS}) and (\ref{DELS}) respectively and we use the following
notation for the differences of counterterms
\begin{equation}
\Delta Q \equiv \delta Q - \widehat{\delta Q} \hspace{1.2cm}
{\rm with} \hspace{1cm} Q = Z_B, Z_W, g^2, \; {\rm etc...}
\end{equation}
The first four equations provide a check for the values of
${\displaystyle \Delta Z_W, \frac{\Delta g^2}{g^2}, a^b_{11}, a^b_3}$
and $a^b_{12}$ that we already obtained in our previous work from the
calculation of the two and three-point functions.  Finally, using
these results and the values of $a^b_8, a^b_9$ and $a^b_{13}$ from
\cite{HR2} we can extract the genuine parameters of the four-point
functions: $ a^b_4, a^b_5, a^b_6, a^b_7$ and $ a^b_{10}$.
By solving the complete linear system of equations, one gets
a unique solution for the bare electroweak chiral parameters
given by:
\begin{eqnarray}
a_0^b &  =  &  \frac{1}{16 \pi^2} \frac{3}{8}
\left( \Delta_\epsilon - \log \frac{\mh^2}{\mu^2} +
\frac{5}{6}\right), \nonumber \\[2mm]
a_1^b & = &  \frac{1}{16 \pi^2} \frac{1}{12}
\left( \Delta_\epsilon - \log \frac{\mh^2}{\mu^2}
+ \frac{5}{6} \right), \nonumber \\[2mm]
a_2^b & = &  \frac{1}{16 \pi^2}
\frac{1}{24} \left( \Delta_\epsilon - \log
\frac{\mh^2}{\mu^2} +
\frac{17}{6} \right), \nonumber \\[2mm]
a_3^b & = & \frac{-1}{16 \pi^2} \frac{1}{24}
\left( \Delta_\epsilon - \log \frac{\mh^2}{\mu^2} +
\frac{17}{6} \right), \nonumber \\[2mm]
a_4^b & = & \frac{-1}{16 \pi^2} \frac{1}{12}
\left( \Delta_\epsilon - \log \frac{\mh^2}{\mu^2} +
\frac{17}{6}\right), \nonumber \\[2mm]
a_5^b & = & \frac{M_W^2}{2 g^2 \mh^2} - \frac{1}{16 \pi^2} \frac{1}{24}
\left( \Delta_\epsilon - \log \frac{\mh^2}{\mu^2} + \frac{79}{3}
- \frac{ 27 \pi}{2 \sqrt{ 3}} \right), \nonumber\\[2mm]
a_{11}^b  & = &
\frac{-1}{16 \pi^2}\frac{1}{24}, \nonumber \\[2mm]
a_6^b & = & a_7^b \; =\; a_8^b \; = \; a_9^b \; = \; a_{10}^b \; = \;
a_{12}^b \; = \; a_{13}^b \; = \; 0. \label{aMH}
\end{eqnarray}

We would like to make some remarks on this result for the chiral
parameters:
\begin{enumerate}
\item First of all, we agree with the $1/\epsilon$ dependence
of the $a_i^b$ parameters that was first calculated by Longhitano
\cite{LON} looking at the divercences of the non-linear sigma model.
We see therefore that the divergences generated with the $\nll$ to
one loop are exactly canceled by the $1/\epsilon$ terms in the
$a_i^b$'s.
\item The values of $a^b_4$ and $a^b_5$ agree with the results
given in \cite{DH}, where the equivalence theorem  was used in
comparing the scattering amplitudes for Goldstone bosons in the SM
\cite{DW} and the EChL. These values, on the
other hand, do not coincide with the values in \cite{EH}
where just contributions from pure Higgs loops were considered.
\item It is important to realize that the matching procedure fixes
completely the values of the bare parameters $a_i^b$ in terms of the
renormalized parameters of the SM.
\item Eqs.(\ref{aMH}) give the complete non-decoupling effects
of a heavy Higgs, that is, the leading logarithmic dependence
on $\mh$ and the next to leading constant contribution
to the electroweak chiral parameters.
The $a_i$'s are accurate up to corrections of the order
$(p/\mh)$ where $p \approx M_Z$ and higher order
corrections in the perturbative expansion.
\item We demonstrate that the $a_i$'s are gauge independent, as
expected.
\item  There is only one effective operator, the one corresponding to
$a_5$, that gets a tree level contribution.
Its expression in terms of renormalized SM parameters
depends on the renormalization prescription that one has
chosen in the standard model, on-shell in our case.
We believe it is important at this point to clarify the relation
among the  $a^b_i$'s that correspond to an on-shell
renormalization of the SM and their corresponding values if a
different renormalization  prescription for the SM is chosen.
We will discuss this point in the following section.
\end{enumerate}

By putting together the results of the two, three and four-point
functions, one also obtains some relations among the counterterms
of the two theories
\begin{eqnarray}
\Delta Z_W & = & \frac{- g^2}{16 \pi^2} \frac{1}{12} \left(
\Delta_\epsilon - \log \frac{\mh^2}{\mu^2} + \frac{5}{6} \right),
\nonumber\\[2mm]
\Delta Z_B & = & \frac{- g'^2}{16 \pi^2} \frac{1}{12} \left(
\Delta_\epsilon - \log \frac{\mh^2}{\mu^2} + \frac{5}{6} \right),
\nonumber\\[2mm]
\Delta \xi_W & = & \Delta Z_W, \hspace{1cm}
\Delta \xi_B  =  \Delta Z_B, \nonumber \\[2mm]
\frac{\Delta g^2}{g^2} & = & \frac{\Delta g'^2}{g'^2}
= 0, \nonumber \\[2mm]
\Delta Z_\phi - 2 \frac{\Delta v}{v} & = & \frac{g^2}{16 \pi^2}
\left[ - \frac{\mh^2}{8 M_W^2} + \frac{3}{4} \left(
\Delta_\epsilon - \log \frac{\mh^2}{\mu^2} + \frac{5}{6} \right)
\right. \nonumber \\[2mm]
& & \left. + \frac{1}{4} \frac{\xi_Z}{c^2}  \left(
\Delta_\epsilon - \log \frac{\xi_Z M_Z^2}{\mu^2} + 1 \right)
+ \frac{1}{2} \xi_W  \left(
\Delta_\epsilon - \log \frac{\xi_W M_W^2}{\mu^2} + 1 \right)
\right]
\end{eqnarray}
These equations give the differences among the renormalization
constants of the SM in the large $\mh$ limit and those in the
EChL, when the on-shell renormalization scheme is chosen in
both theories. They are obtained here as a constraint imposed
by the matching; one can also calculate them from the explicit
expressions of the on-shell counterterms of the two theories
and verify that these relations are indeed satisfied.

\section{ Dependence of the chiral parameters on the renormalization
of the SM}

In the previous section, we have given the values of the bare
electroweak chiral parameters (\ref{aMH}) when the SM is renormalized
in the on-shell scheme.  We would like now to discuss how these bare
parameters are changed when a different renormalization scheme is
chosen for the SM.

We have seen that the chiral parameter $a^b_5$ is the only one
that gets a tree level contribution when a heavy Higgs is integrated out,
and therefore it is the only one whose expression in terms of the
SM renormalized parameters will depend on the renormalization
prescription of the SM.
By using eqs.(\ref{CORE}) and (\ref{aMH}), one can rewrite $a^b_5$ in terms
of the on-shell renormalized scalar self-coupling $\lambda$:
\begin{equation}
a^b_5 = \frac{1}{16 \lambda} -
\frac{1}{16 \pi^2} \frac{1}{24}
\left( \hat{\Delta}_\epsilon + \frac{79}{3}
- \frac{ 27 \pi}{2 \sqrt{ 3}} \right). \label{a5LR}
\end{equation}
The easiest way to connect with a new renormalization scheme for
the SM is to write down $a^b_5$ in terms of the bare scalar
self-coupling $\lambda^b$. To this end, we first write down
the relation\footnote{The contribution from the
renormalization of $g$ is of higher order in the large $M_H$
expansion than the contributions
 from the renormalization of $M_H$ and $M_W$ and will be ignored
here.} between the renormalized self-coupling $\lambda$
and the bare $\lambda^b$
\begin{equation}
\lambda = \lambda^b \left( 1 + \frac{\delta \mh^2}{\mh^2} -
 \frac{\delta M_W^2}{M_W^2} \right)
\end{equation}
By substituing the values of the mass counterterms given in
(\ref{SMCL}) in this equation, we get the relation among the
bare and the renormalized self-coupling in the on-shell scheme
\begin{equation}
\lambda = \lambda^b \left[ 1 +
\frac{1}{16 \pi^2} \lambda^b \left( -12  \hat{\Delta}_\epsilon
-25 + \frac{9 \pi}{\sqrt 3} \right) \right]
\end{equation}
Now, by plugging the last equation into eq.(\ref{a5LR})
one finds the value of $a^b_5$ in terms of $\lambda^b$
\begin{equation}
a^b_5 = \frac{1}{16 \lambda^b} +
\frac{1}{16 \pi^2} \frac{1}{24}
\left( 17 \hat{\Delta}_\epsilon + \frac{67}{6}\right).\label{a5LB}
\end{equation}

In order to connect with a different renormalization prescription
we simply substitute $\lambda^b$ in eq.(\ref{a5LB}) by its
corresponding definition in terms of the new renormalized
self-coupling.
For instance, if one chooses the
$\msb$ scheme, where the scalar self-coupling is
defined by
\begin{equation}
\lambda^b = \lambda_\msb \left[ 1 + \frac{1}{16 \pi^2} \;
 \lambda_\msb \; 12 \; \hat{\Delta}_\epsilon \right]
\end{equation}
then $a^b_5$ in terms of the renormalized self-couplig
$\lambda_\msb$ is given by
\begin{equation}
a^b_5 = \frac{1}{16 \lambda_\msb} -
\frac{1}{16 \pi^2} \frac{1}{24}
\left( \hat{\Delta}_\epsilon - \frac{67}{6}\right).
\end{equation}
and for the rest of chiral parameters one gets exactly the
same values as in eq.(\ref{aMH}).

Another example is the renormalization prescription chosen by
Gasser and Leutwyler in \cite{GL1}.
Their prescription for the renormalized coupling is given by
\footnote{In ref.\cite{GL1} $g_r$ is what we call here $\lambda_{\rm GL}$.
The mass apearing in $\hat{\Delta}_\epsilon$
is not the physical mass of the Higgs, but what they call $M_r$.
The renormalized mass $M_r$ was fixed in \cite{GL1} such that
$M_r^2 = 2 \lambda_{\rm GL} v^2$.}
\begin{equation}
\lambda^b = \lambda_{\rm GL} \left[ 1 + \frac{1}{16 \pi^2} \;
 \lambda_{\rm GL} \; 12 \; ( \hat{\Delta}_\epsilon + 1) \right]
\end{equation}
and the expression of $a^b_5$ in terms of the renormalized
self-coupling $\lambda_{\rm GL}$ is therefore
\begin{equation}
a^b_5 = \frac{1}{16 \lambda_{\rm GL}} -
\frac{1}{16 \pi^2} \frac{1}{24}
\left( \hat{\Delta}_\epsilon + \frac{41}{6}\right).\label{a5GL}
\end{equation}
and the rest of parameters remain again the same.

For comparison, it is interesting to translate the results
of eqs.(\ref{aMH},\ref{a5GL}) to the usual notation in chiral
perturbation theory
\begin{eqnarray}
L^b_1 & = & \frac{l^b_1}{4} = a^b_5 =
\frac{1}{16 \lambda_{\rm GL}} -
\frac{1}{16 \pi^2} \frac{1}{24}
\left( \hat{\Delta}_\epsilon + \frac{41}{6}\right).
\nonumber \\[2mm]
L^b_2 & = & \frac{l^b_2}{4} = a^b_4 =
\frac{- 1}{16 \pi^2} \frac{1}{12}
\left( \hat{\Delta}_\epsilon  +
\frac{17}{6}\right), \nonumber \\[2mm]
L^b_9 & = & - \frac{ l^b_6}{2} = a^b_3 - a^b_2 =
\frac{ - 1}{16 \pi^2} \frac{1}{12}
\left( \hat{\Delta}_\epsilon
+ \frac{17}{6} \right), \nonumber \\[2mm]
L^b_{10} & = & l^b_5 = a^b_1 =
 \frac{1}{16 \pi^2} \frac{1}{12}
\left( \hat{\Delta}_\epsilon
+ \frac{5}{6} \right). \label{LiB}
\end{eqnarray}

These chiral parameters agree with the values found by Gasser
and Leutwyler in \cite{GL1} (see appendix B of this reference)
and in \cite{NS}.  We find it a quite remarkable agreement since
they used functional methods to integrate out the Higgs particle
in a linear sigma model where the gauge fields were considered
as external sources and they used as well different techniques
to study the large $\mh$ limit.  This confirms the fact already
mentioned that the contributions to the effective operators of
dimension four that come from mixed gauge-scalar loops are
subleading, in the large $\mh$ limit, compared to the pure
scalar loops contributions.  However, this is not the case for
the dimension two custodial breaking operator $a_0$ that gets
contributions from mixed gauge-scalar loops \cite{HR2}.

To conclude this section, we emphasize once more that the bare chiral
parameters for a given underlying theory, the SM in our case, must be
computed with a choice for the renormalization prescription of this
theory.  The explicit expression of the chiral parameters will vary
from one prescription to another, but the numerical value remains the
same, and the connection between different prescriptions  can be
clearly and easily established.

\section{Renormalization of the effective theory}

In this section we briefly describe the renormalization procedure in
the effective theory.  Given the effective Lagrangian of
eq.(\ref{ECL}), the first step is to redefine the fields and
parameters of the Lagrangian according to eq.(\ref{RET}).  It
introduces, at a formal level, the set of counterterms of the
effective theory $\widehat{\delta Z}_i, \widehat{\delta g}$, etc,
that need to be computed once a particular renormalization
prescription scheme is chosen. We fix here the on-shell
renormalization scheme as we did in the case of the SM. For practical
reasons we prefer to choose the renormalization conditions as in
reference \cite{HO}, which are the most commonly used for LEP
physics. In terms of the renormalized selfenergies these
renormalization conditions read as follows
\begin{equation}
\widehat{\Sigma}^{\rm R}_{W} (M_W^2) = 0,\hspace{3mm}
\widehat{\Sigma}^{\rm R}_{Z} (M_Z^2) = 0,\hspace{3mm}
\widehat{\Sigma}^{\prime \; \rm R}_{\gamma} (0) = 0,\hspace{3mm}
\widehat{\Sigma}^{\rm R}_{\gamma Z} (0) = 0. \label{RCE}
\end{equation}
The renormalized self energies are computed in the effective theory
as usual, namely, by adding all the contributions from the one
loop diagrams and from the counterterms. We get the following
expressions\footnote{Notice that in our rotation defining the physical
gauge fields, the terms in $s$ have different sign than in reference
\cite{HO}}:
\begin{eqnarray}
\widehat{\Sigma}^{\rm R}_{\gamma} (q^2) & = &
\widehat{\Sigma}^{\rm L}_{\gamma} (q^2) +
\left( s^2 \widehat{\delta Z_W} + c^2 \widehat{\delta Z_B}
\right) q^2 + s^2 g^2 (a^b_8 - 2 a^b_1) q^2.
\nonumber\\[2mm]
\widehat{\Sigma}^{\rm R}_{W} (q^2) & = &
\widehat{\Sigma}^{\rm L}_{W} (q^2) +
\widehat{\delta Z_W} \left( q^2 - M_W^2 \right)
- \widehat{\delta M_W^2}. \nonumber\\[2mm]
\widehat{\Sigma}^{\rm R}_{Z} (q^2) & = &
\widehat{\Sigma}^{\rm L}_{Z} (q^2) +
\left( c^2 \widehat{\delta Z_W} + s^2 \widehat{\delta Z_B}
\right) ( q^2 - M_Z^2) - \widehat{\delta M_Z^2} \nonumber\\
& &
+ 2 g'^2 a_0^b M_Z^2 + \left( 2 s^2 g^2  a^b_1 + c^2 g^2 a_8^b
 + (g^2 + g^{\prime 2}) a^b_{13} \right) q^2.
\nonumber\\[2mm]
\widehat{\Sigma}^{\rm R}_{\gamma Z} (q^2) & = &
\widehat{\Sigma}^{\rm L}_{\gamma Z} (q^2) +
s c \left( \widehat{\delta Z_W} - \widehat{\delta Z_B}
\right) q^2 - s c\; M_Z^2 \left( \frac{\widehat{\delta g'}}{g'} -
\frac{\widehat{\delta g}}{g} \right) \nonumber\\
& & + \left( s c g^2 a^b_8 - (c^2 - s^2) g g' a^b_1 \right) q^2.
\label{JQL}
\end{eqnarray}
where
\begin{eqnarray}
\widehat{\delta M_W^2} & = & M_W^2 \left(
\widehat{\delta Z_\Phi} - 2 \frac{\widehat{\delta g}}{g}
- 2 \frac{\widehat{\delta v}}{v} - \widehat{\delta Z_W}
\right), \nonumber\\[2mm]
\widehat{\delta M_Z^2} & = & M_Z^2 \left(
\widehat{\delta Z_\Phi} - 2 c^2 \frac{\widehat{\delta g}}{g}
- 2 s^2 \frac{\widehat{\delta g'}}{g'}
- 2 \frac{\widehat{\delta v}}{v} -
c^2 \widehat{\delta Z_W} - s^2 \widehat{\delta Z_B}
\right),\nonumber\\[2mm]
M_W^2 & = & g^2 v^2 / 4, \nonumber\\[2mm]
M_Z^2 & = & (g^2 + g'^2) v^2 / 4,
\label{masas}
\end{eqnarray}
and the superscripts R and L denote renormalized and EChL
loops respectively.

{}From eq.(\ref{masas}) the following relation among the
$W$ and $Z$ mass counterterms is obtained
\begin{equation}
\frac{\widehat{\delta M_Z^2}}{M_Z^2} -
\frac{\widehat{\delta M_W^2}}{M_W^2} =
 2 s^2 \frac{\widehat{\delta g}}{g} +
 2 c^2 \frac{\widehat{\delta g'}}{g'}
+ s^2 \left(\widehat{\delta Z_W} - \widehat{\delta Z_B}
\right)
\end{equation}
Finally, by requiring these renormalized self energies to fulfill
eqs.(\ref{RCE}) and taking into account that the $U(1)_{\rm Y}$
Ward identity implies $\widehat{\delta g'} = 0$
one gets the following results for the values of the counterterms
in terms of the unrenormalized selfenergies of the effective
theory and the bare $a_i$'s:
\begin{eqnarray}
\widehat{\delta M_W^2} & = &
\widehat{\Sigma}^{\rm L}_W (M_W^2), \nonumber\\[2mm]
\widehat{\delta M_Z^2} & = &
\widehat{\Sigma}^{\rm L}_{Z} (M_Z^2) + M_Z^2
\left( 2 g'^2 a_0^b + 2 s^2 g^2  a^b_1 + c^2 g^2 a_8^b
 + (g^2 + g^{\prime 2}) a^b_{13} \right), \nonumber\\[2mm]
\frac{\widehat{\delta g}}{g} & = &
\frac{- 1}{s c} \frac{\widehat{\Sigma}^{\rm L}_{ \gamma Z}(0)}
{M_Z^2}, \nonumber\\[2mm]
\frac{\widehat{\delta g'}}{g'} & = & 0, \nonumber\\[2mm]
\widehat{\delta Z_W} & = &
\frac{c^2}{s^2} \left(\frac{\widehat{\Sigma}^{\rm L}_{Z}(M_Z^2)}
{M_Z^2} - \frac{\widehat{\Sigma}^{\rm L}_{W}(M_W^2)}
{M_W^2} \right) + 2 \frac{c}{s} \frac{
\widehat{\Sigma}^{\rm L}_{ \gamma Z}(0)}{M_Z^2} -
\widehat{\Sigma}^{\prime \;\rm L}_\gamma (0) \nonumber\\
& & + 2 g^2 a^b_0 + 2 g^2 a^b_1 +
\frac{c^2 - s^2}{s^2} g^2 a^b_8 + \frac{c^2}{s^2}
(g^2 + g'^2) a^b_{13}, \nonumber\\[2mm]
\widehat{\delta Z_B} & = &
 \frac{\widehat{\Sigma}^{\rm L}_{W}(M_W^2)}{M_W^2} -
\frac{ \widehat{\Sigma}^{\rm L}_{Z}(M_Z^2)}{M_Z^2} -
2 \frac{s}{c} \frac{\widehat{\Sigma}^{\rm L}_{ \gamma Z}(0)}{M_Z^2} -
\widehat{\Sigma}^{\prime \rm L}_\gamma (0) \nonumber\\
& & - \left( 2 g'^2 a^b_0 + g^2 a^b_8 +
(g^2 + g'^2) a^b_{13} \right). \label{ECT}
\end{eqnarray}

Now that we have at hand eqs.(\ref{ECT}) the only parameters of the
theory that still need to be renormalized are the electroweak chiral
parameters $a_i$. The following formal redefinition of the chiral
parameters has already been introduced
\begin{equation}
a^b_i = a_i (\mu) + \delta a_i. \label{NSQ}
\end{equation}
The divergent part of the $a_i^b$ parameters, or equivalently the
divergent part of the counterterms $\delta a_i$, are fixed by
the symmetries of the effective theory and since the work of
Longhitano \cite{LON} they are known to be
\begin{eqnarray}
& & \delta a_0 |_{div} = \frac{1}{16 \pi^2} \frac{3}{8} \Delta_\epsilon ,
\hspace{2cm}\delta a_1 |_{div} = \frac{1}{16 \pi^2} \frac{1}{12}
\Delta_\epsilon , \nonumber \\[2mm]
& & \delta a_2 |_{div} = \frac{1}{16 \pi^2} \frac{1}{24} \Delta_\epsilon ,
\hspace{2cm}\delta a_3 |_{div} = \frac{- 1}{16 \pi^2} \frac{1}{24}
\Delta_\epsilon , \nonumber \\[2mm]
& & \delta a_4 |_{div} = \frac{- 1}{16 \pi^2} \frac{1}{12} \Delta_\epsilon ,
\hspace{2cm}\delta a_5 |_{div} = \frac{- 1}{16 \pi^2} \frac{1}{24}
\Delta_\epsilon , \nonumber \\[2mm]
& & \delta a_i |_{div} = 0, \hspace{5mm} i = 6,....13. \label{PP}
\end{eqnarray}
These universal divergent contributions to the chiral bare
parameters imply in turn universal renormalization group
equations for the renormalized parameters
\begin{eqnarray}
& &a_0(\mu)  =	a_0(\mu ') + \frac{1}{16 \pi^2}
\frac{3}{8} \log\frac{\mu^2}{\mu '^2}, \hspace{1cm}
a_1(\mu)  =  a_1(\mu ') + \frac{1}{16 \pi^2}
\frac{1}{12} \log\frac{\mu^2}{\mu '^2}, \nonumber\\[2mm]
& &a_2(\mu)  =	a_2(\mu ') + \frac{1}{16 \pi^2}
\frac{1}{24} \log\frac{\mu^2}{\mu '^2}, \hspace{1cm}
a_3(\mu)  =  a_3(\mu ') - \frac{1}{16 \pi^2}
\frac{1}{24} \log\frac{\mu^2}{\mu '^2}, \nonumber\\[2mm]
& &a_4(\mu)  =	a_4(\mu ') - \frac{1}{16 \pi^2}
\frac{1}{12} \log\frac{\mu^2}{\mu '^2}, \hspace{1cm}
a_5(\mu)  =  a_5(\mu ') - \frac{1}{16 \pi^2}
\frac{1}{24} \log\frac{\mu^2}{\mu '^2}, \nonumber\\[2mm]
& &a_i(\mu) = a_i(\mu') ; \hspace{3mm} i = 6,...13.
\end{eqnarray}

The value of the bare chiral parameters $a^b_i$, on the other hand,
is completely determined by the matching procedure in terms of the
renormalized parameters of the underlying physics that has been
integrated out, as we have seen for the particular case of a heavy
Higgs.	However, for a given $a^b_i$, we still have to choose how to
separate the finite part into the renormalized $a_i(\mu)$ and the
counterterm $\delta a_i$ in eq.(\ref{NSQ}) such that their sum gives
$a^b_i$.  This second renormalization scheme concerns only to the
effective theory.  Therefore, in using a set of renormalized
parameters $a_i(\mu)$ for a particular underlying theory, one must
specify, in addition, how the finite parts of the counterterms in
eq.(\ref{NSQ}) have been fixed.

In the case of the SM, where a heavy Higgs has been integrated
out to one loop, the bare chiral parameters are given in
eq.(\ref{aMH}). They correspond to the on-shell renormalization
of the underlying SM. Now, in order to present the corresponding
renormalized parameters we have to fix the finite parts of the
counterterms. For instance, if we fix the counterterms to include
just the $\Delta_\epsilon$ terms as in eq.(\ref{PP}),
the renormalized chiral parameters for the SM with a heavy Higgs
are\footnote{This particular renormalization of the chiral parameters
was chosen in our previous work, where we called it $\msb$.}:
\begin{equation}
\begin{array}{ll}
a_0 (\mu)  =  {\displaystyle  \frac{1}{16 \pi^2}
\frac{3}{8}\left( \frac{5}{6} - \log\frac{M_H^2}{\mu^2} \right)},&
\hspace{-1cm}a_3 (\mu) =  {\displaystyle \frac{-1}{16 \pi^2}
\frac{1}{24}
\left( \frac{17}{6} - \log\frac{M_H^2}{\mu^2} \right)},\\[6mm]
a_1 (\mu)  =  {\displaystyle \frac{1}{16 \pi^2}  \frac{1}{12}
\left( \frac{5}{6} - \log\frac{M_H^2}{\mu^2} \right)}, &
\hspace{-1cm}a_4 (\mu) =  {\displaystyle \frac{- 1}{16 \pi^2}
\frac{1}{12}
\left( \frac{17}{6} - \log\frac{M_H^2}{\mu^2} \right)},\\[6mm]
a_2 (\mu)  =  {\displaystyle \frac{1}{16 \pi^2} \frac{1}{24}
\left( \frac{17}{6} - \log\frac{M_H^2}{\mu^2} \right)}, &
\hspace{-1cm}a_{11} (\mu)  = {\displaystyle
\frac{-1}{16 \pi^2}\frac{1}{24}},\\[6mm]
a_5 (\mu)  =  {\displaystyle \frac{v^2}{8 \mh^2} -
 \frac{1}{16 \pi^2}  \frac{1}{24}
\left( \frac{79}{3} - \frac{ 27 \pi}{2 \sqrt{3}}
 - \log\frac{M_H^2}{\mu^2} \right)}, \\[6mm]
a_i (\mu) = 0, \hspace{3mm} i = 6,7,8,9,10,12,13.
\end{array} \label{aR}
\end{equation}

Another example is the renormalization scheme chosen by Gasser and
Leutwyler for the linear $O(N)$ sigma model in ref.\cite{GL1}.
The values of the bare parameters for their choice of the renormalization
scheme of the underlying sigma model were given in eqs.(\ref{LiB}).
Now, they fix instead the chiral counterterms to the following values:
\begin{eqnarray}
\delta L_1 & = & \frac{\delta l_1}{4} = \delta a_5^{\rm GL} =
\frac{1}{16 \pi^2} \frac{- 1}{24} ( \Delta_\epsilon + 1),
\nonumber \\[2mm]
\delta L_2 & = & \frac{\delta l_2}{4} = \delta a_4^{\rm GL} =
\frac{1}{16 \pi^2} \frac{- 1}{12} ( \Delta_\epsilon + 1),
\nonumber \\[2mm]
\delta L_9 & = & -  \frac{\delta l_6}{2} = \delta a_3^{\rm GL}
- \delta a_2^{\rm GL} = \frac{1}{16 \pi^2} \frac{- 1}{12}
( \Delta_\epsilon + 1), \nonumber \\[2mm]
\delta L_{10} & = & \delta l_5 = \delta a_1^{\rm GL} =
\frac{1}{16 \pi^2} \frac{1}{12} ( \Delta_\epsilon + 1).
\end{eqnarray}

Consequently, they get the following renormalized chiral
parameters for the sigma model\footnote{Here we call
$M_{\rm GL}$ the renormalized mass $M_r$ of \cite{GL1}.}
\begin{eqnarray}
L_1(\mu) & = & \frac{l_1(\mu)}{4} = a_5^{\rm GL}(\mu) =
\frac{1}{16 \lambda_{\rm GL}} -
\frac{1}{16 \pi^2} \frac{1}{24}
\left( \frac{35}{6} - \log\frac{M^2_{\rm GL}}{\mu^2} \right),
\nonumber \\[2mm]
L_2(\mu) & = & \frac{l_2(\mu)}{4} = a_4^{\rm GL}(\mu) =
\frac{- 1}{16 \pi^2} \frac{1}{12}
\left( \frac{11}{6} - \log\frac{M^2_{\rm GL}}{\mu^2}
\right), \nonumber \\[2mm]
L_9(\mu) & = & - \frac{ l_6(\mu)}{2} = a_3^{\rm GL}(\mu)
- a_2^{\rm GL}(\mu) =
\frac{ - 1}{16 \pi^2} \frac{1}{12}
\left( \frac{11}{6} - \log\frac{M^2_{\rm GL}}{\mu^2}
\right), \nonumber \\[2mm]
L_{10}(\mu) & = &  l_5(\mu) = a_1^{\rm GL}(\mu) =
 \frac{1}{16 \pi^2} \frac{1}{12}
\left( - \frac{1}{6} - \log\frac{M^2_{\rm GL}}{\mu^2}
\right).
\end{eqnarray}

\section{Calculating observables with the EChL}

In this section we will show, as an example, the explicit calculation
of the radiative corrections to $\Delta \rho$ and $\Delta r$ within
the electroweak chiral Lagrangian approach. These observables are
defined in the effective theory in terms of the renormalized
self-energies in the same way as in the fundamental SM, namely:
\begin{eqnarray}
\Delta \rho & \equiv & \frac{\widehat{\Sigma}^{\rm R}_Z (0)}{M_Z^2}
-  \frac{\widehat{\Sigma}^{\rm R}_W (0)}{M_W^2}, \nonumber\\[2mm]
\Delta r & \equiv &  \frac{\widehat{\Sigma}^{\rm R}_W (0)}{M_W^2}
+ {\rm (vertex + box)}, \label{ROR}
\end{eqnarray}
where
\begin{displaymath}
{\rm (vertex + box)} \equiv \frac{g^2}{16 \pi^2} \left(
6 + \frac{ 7 - 4 s^2}{2 s^2} \log c^2 \right),
\end{displaymath}
and the renormalized self-energies can be computed as we have
explained in section 6.

Once a renormalization scheme has been chosen, one can always express
$\Delta \rho$ and $\Delta r$ in terms of unrenormalized self-energies
and the $a_i^b$'s. For instance, in the on-shell scheme
given by the conditions of eq.(\ref{RCE}), one gets the particular
values of the counterterms given in eqs.(\ref{ECT}). Next, by plugging
these counterterms into eqs.(\ref{JQL}), one obtains the renormalized
self-energies in terms of the unrenormalized ones and the
$a_i^b$'s. Finally, by substituing these formulas into
eqs.(\ref{ROR}) the following expressions for $\Delta \rho$ and
$ \Delta r$ in the on-shell scheme are found
\begin{eqnarray}
\Delta \rho & = & \frac{\widehat{\Sigma}^{\rm L}_Z (0)}{M_Z^2}
-  \frac{\widehat{\Sigma}^{\rm L}_W (0)}{M_W^2}
+ \frac{2 s}{c} \frac{\widehat{\Sigma}^{\rm L}_{\gamma Z} (0)}{M_Z^2}
+ 2 g'^2 a^b_0, \nonumber \\[2mm]
\Delta r & = & \frac{\widehat{\Sigma}^{\rm L}_W (0) -
\widehat{\Sigma}^{\rm L}_W (M_W^2)}{M_W^2} +
\widehat{\Sigma}'^{\rm  L}_\gamma (0) + \frac{c^2}{s^2}
\left[ \frac{\widehat{\Sigma}^{\rm L}_W (M_W^2)}{M_W^2}
- \frac{\widehat{\Sigma}^{\rm L}_Z (M_Z^2)}{M_Z^2}
-  \frac{2 s}{c} \frac{\widehat{\Sigma}^{\rm L}_{\gamma Z} (0)}
{M_Z^2} \right] \nonumber\\[2mm]
& & - 2 g^2 a^b_0 + \frac{s^2 - c^2}{s^2}
g^2  (a^b_8 + a^b_{13} ) - 2 g^2 (a^b_1 + a^b_{13}) +
 {\rm (vertex + box)}.
\end{eqnarray}

The explicit computation of the bosonic loop contributions in the
effective theory, as well as the contributions from just
$a^b_0$ and $a^b_1$ were found in \cite{DEH,EH}.
We present here the complete result
\begin{eqnarray}
\Delta \rho & = & \frac{g^2}{16 \pi^2} \left[
 \frac{3}{4} \frac{s^2}{c^2} \left( - \Delta_\epsilon
+ \log \frac{M_W^2}{\mu^2} \right) + h(M_W^2,M_Z^2)
 \right]
+ 2 g'^2 a^b_0, \nonumber \\[2mm]
\Delta r & = & \frac{g^2}{16 \pi^2} \left[
\frac{11}{12} \left( \Delta_\epsilon - \log \frac{M_W^2}{\mu^2}
\right) + f(M_W^2,M_Z^2) \right] \nonumber\\[2mm]
& & - 2 g^2 a^b_0 + \frac{s^2 - c^2}{s^2}
g^2  (a^b_8 + a^b_{13} ) - 2 g^2 (a^b_1 + a^b_{13}). \label{RORAB}
\end{eqnarray}
where
\begin{eqnarray}
h(M_W^2,M_Z^2) & = &  \frac{1}{c^2} \log c^2 \left( \frac{17}{4 s^2} -
7 + 2 s^2 \right) + \frac{17}{4} - \frac{5}{8} \frac{s^2}{c^2}
\nonumber\\[2mm]
f(M_W^2,M_Z^2) & = & \log c^2 \left( \frac{5}{c^2} -1 +
\frac{3 c^2}{s^2} - \frac{17}{4 s^2 c^2} \right)
- s^2 (3 + 4 c^2) F(M_Z^2,M_W,M_W)\nonumber \\
& & + I_2(c^2)(1 - \frac{c^2}{s^2}) + \frac{c^2}{s^2} I_1(c^2) +
\frac{1}{8 c^2} ( 43 s^2 - 38) \nonumber \\
& & + \frac{1}{18} (154 s^2 - 166 c^2) + \frac{1}{4 c^2} +
\frac{1}{6} + \Delta \alpha +
\left( 6 + \frac{7 - 4 s^2} {2 s^2} \log c^2 \right).
\end{eqnarray}
and $F$, $I_1$, $I_2$ and $\Delta \alpha$ can be found in \cite{MS}.
In eqs.(\ref{RORAB}) there are apparently a divergent term and a $\mu$-scale
dependent term. However, when one redefines the bare effective chiral
parameters as usual, $a^b_i = a_i(\mu) + \delta a_i$, it can be easily
seen that the divergent terms are cancelled by the divergent parts
of the $\delta a_i$ and the $\mu$-scale dependence is cancelled by the
scale dependence of the $a_i(\mu)$. The observables $\Delta \rho$ and
$\Delta r$ turn out to be finite and scale and renormalization
prescription independent, as it must be. In particular, if we set the
substraction scheme for the chiral counterterms
to include just the $\Delta_\epsilon$ terms as in eq.(\ref{PP}),
the following expressions for $\Delta \rho$ and $\Delta r$
in terms of renormalized chiral parameters are obtained
\begin{eqnarray}
\Delta \rho & = & \frac{g^2}{16 \pi^2} \left[
 \frac{3}{4} \frac{s^2}{c^2} \log \frac{M_W^2}{\mu^2}
+ h(M_W^2, M_Z^2) \right]
+ 2 g'^2 a_0(\mu), \nonumber \\[2mm]
\Delta r & = & \frac{g^2}{16 \pi^2} \left[
 - \frac{11}{12}
\log \frac{M_W^2}{\mu^2} + f(M_W^2,M_Z^2) \right] \nonumber\\[2mm]
& & - 2 g^2 a_0(\mu) + \frac{s^2 - c^2}{s^2}
g^2  (a_8 + a_{13} ) - 2 g^2 (a_1(\mu) + a_{13}). \label{RRF}
\end{eqnarray}

Equations (\ref{RRF}) are general and can be applied to any
underlying physics for the symmetry breaking sector.
If we want to recover the values of $\Delta \rho$ and $\Delta r$
in the particular case of the SM with a heavy Higgs,
one just has to substitute the values of the chiral parameters
in eqs.(\ref{aR}) into eqs.(\ref{RRF}) to obtain
\begin{eqnarray}
\Delta \rho & = & \frac{g^2}{16 \pi^2} \left[
 - \frac{3}{4} \frac{s^2}{c^2} \left(
\log \frac{M_H^2}{M_W^2} - \frac{5}{6} \right)
+ h(M_W^2, M_Z^2) \right], \nonumber \\[2mm]
\Delta r & = & \frac{g^2}{16 \pi^2} \left[
 \frac{11}{12}
\left(\log \frac{M_H^2}{M_W^2} - \frac{5}{6} \right)
 + f(M_W^2,M_Z^2) \right].
\end{eqnarray}
which agrees with the result given in \cite{MS}.

One can similarly obtain the heavy Higgs contributions to other
relevant observables in electroweak phenomenology.

\section{Conclusions}

Given the present situation of remarkable improvement in electroweak
precision measurements and the encouraging prospects for the future,
we believe it is now imperative a good undertanding of the SM
Higgs boson radiative corrections. A heavy Higgs boson do not
decouple from the low energy electroweak observables and therefore
will leave some trace on them that could be observed with the
future precision measurements.

In this paper, we have calculated the complete non-decoupling
effects of the SM Higgs boson to one loop. They consist of the
already known leading logarithmic Higgs mass dependent effects
and the next to leading constant terms. Both effects are
relevant and for not too large $\mh$ values can be of
comparable magnitude.

We have classified these non-decoupling effects using the EChL
approach, an effective field theory that respects the SM
symmetries at low energies. Within this approach, the
non-decoupling effects of a heavy Higgs boson are represented,
at energies below the Higgs mass, by a subset of gauge invariant
effective operators of the EChL. We have calculated in this work
the values of the parameters of these chiral operators that
represent the SM Higgs.  It is our main result and is given in
eq.(\ref{aMH}).  We get just seven non-vanishing relevant
parameters summarizing the whole set of non-decoupling heavy
Higgs effects to one loop.
We have also discussed in detail in this work the relation
between the renormalization of both the SM and the effective
theory, with special emphasis in the on-shell scheme, that
has been our particular choice here.

In conclusion, we believe that the approach followed in this
work is interesting and useful because it provides a gauge
invariant way of separating the non-decoupling Higgs boson
effects from the rest of the EW radiative corrections and,
on the other hand, it is a general framework in which one can
analyze the low energy effects not only of an SM heavy Higgs,
but of alternative strongly interacting syummetry breaking
scenarios. The EChL that parametrizes the SM Higgs will
then serve as a fundamental reference model.

\section*{Acknowledgements}

We are indebted to C.P. Mart\'{\i}n for his valuable help
with some technical aspects of this work and his interesting comments.
We thank A. Dobado for many useful conversations and for
reading the manuscript. We appreciate the critical reading
of the manuscript done by S. Peris.
We would like to thank also the SLAC theory group and the CERN
Th-division for their hospitality.
M.J.H. acknowledges finantial support from Consejer\'{\i}a de
Educaci\'on de la Comunidad de Madrid during her stay at SLAC.
This work has been partially supported by the spanish
Ministerio de Educaci\'on y Ciencia under project
CICYT AEN93-0673.

\newpage
\setcounter{equation}{0}
\renewcommand{\theequation}{A.\arabic{equation}}
\section*{Appendix A}

In this appendix, we give an explicit example of the
large-$\mh$ techinques used in the calculation of
the loop integrals.

We start by giving the m-theorem of G. Giavarini, C. P. Martin and
F. Ruiz Ruiz \cite{CAR} (reduced to
the case of 1-loop integrals), which gives sufficient
conditions for a loop integral to vanish in the large-m
limit.
Consider an integral of the form
\begin{equation}
I(p,m) = m^\beta \int d^4 k \; \frac{M(k)}{\prod_i (l_i^2
+ m_i^2)^{n_i}} ,
\end{equation}
where
\begin{eqnarray}
l_i & = & k + \sum_{j=1}^{E} b_{ij} p_j \nonumber\\[2mm]
m_i & = & 0 \;\; {\rm or} \;\; m
\end{eqnarray}
and $M(k)$ is a monomial in the components of $k$. $\beta$ denote an
arbitrary real number.
The external momenta $p_1,....,p_E$ lay in a bounded
subdomain of $R^4$. Let $d$ be the mass dimension of $I(p,m)$ and $\omega$
the minimum of zero and the infrared degree of $I(p,m)$
at zero external momenta. Then\\
{\bf m-theorem}: If the integral $I(p,m)$ is both UV and IR
convergent by power counting at non-exceptional external momenta
and $d - \omega < 0 $, then $I(p,m)$ goes to zero when $m$ goes
to infinity.

As an example, consider the $(H-\phi)$ loop correction to the
$W$ self-energy given in fig.(2):
\begin{equation}
\frac{e^2}{4 s^2}  \mu ^{4 - D} \int
\frac{d^D k}{(2 \pi)^D}
\frac{(2 k + q)_\mu (2 k + q)_\nu}{[k^2 - \xi M_W^2] [(k+q)^2 - M_H^2]}
\end{equation}
where $q$ is the $W$ external momentum and $D$ denotes the space-time
dimension in dimensional regularization.

Let's work out explicitly the large $\mh$ expansion of the most
divergent part of this correction which comes from the integral
\begin{equation}
I_{\mu \nu} = \mu^{4 - D} \int
\frac{d^D k}{(2 \pi)^D}
\frac{k_\mu \; k_\nu}{[k^2 - \xi M_W^2] [(k+q)^2 - M_H^2]}
\end{equation}
The superficial degree of UV divergence of $I_{\mu \nu}$ is 2 at $D=4$.
The first step is to rearrange {\sl algebraically}
the denominator that includes a light mass
\begin{equation}
\frac{1}{k^2 -\xi M_W^2} = \frac{1}{k^2} +
\frac{\xi M^2_W}{k^2 (k^2 - \xi M_W^2)}
\end{equation}
so that one gets a term in the integral with the same degree of UV
divergence but where the light mass is no more in the denominator,
and a second term with still a light mass dependence but with the
degree of UV divergence lowered by two.
Applying this algebraic rearrangement as many times as necessary
until the last term gives a convergent integral (twice in the case
of $I_{\mu \nu}$), one gets
\begin{eqnarray}
I_{\mu \nu} &=& \mu^{4 - D} \int \frac{d^D k}{(2 \pi)^D}
\left[ \frac{k_\mu k_\nu}{k^2 [(k+q)^2 - M_H^2]} +
\frac{ \xi M_W^2 k_\mu k_\nu}{k^4 [(k+q)^2 - M_H^2]} +
\frac{\xi^2 M_W^4 k_\mu k_\nu}{ k^4 [k^2 - \xi M_W^2] [(k+q)^2
- M_H^2]} \right] \nonumber \\[2mm]
&=& A_{\mu \nu} + B_{\mu \nu} + C_{\mu \nu}
\end{eqnarray}
Now the last term $C_{\mu \nu}$ is already UV and IR convergent.
One can apply the Lebesgue dominated convergence theorem
to see that the heavy Higgs mass limit can be safely taken
inside the integral in this term, and therefore,
\begin{displaymath}
C_{\mu \nu} \rightarrow 0 \hspace{8mm} {\rm when} \hspace{8mm}
\mh \rightarrow \infty
\end{displaymath}

Let's develope now the term $A_{\mu \nu}$, (the other term
$B_{\mu \nu}$  can be evaluated using the same procedure).
We rewrite the denominators in $A_{\mu \nu}$, using again an
algebraic identity
\begin{equation}
\frac{1}{(k+q)^2 - M_H^2} =
\frac{1}{k^2 - M_H^2} -
\frac{2 k q + q^2}{[k^2 - M_H^2] [(k+q)^2 - M_H^2]}
\end{equation}
One has to apply this identitity as many times as neccessary
until one gets an integral that fulfills the conditions
of the m-theorem.
Using this identity three times in $A_{\mu \nu}$
one get's
\begin{eqnarray}
A_{\mu \nu} &=& \mu^{4-D} \int \frac{d^D k}{(2 \pi)^D}
\left[ \frac{k_\mu k_\nu}{k^2 [k^2 - M_H^2]} -
\frac{k_\mu k_\nu (2 k q + q^2)}{k^2 [k^2 - M_H^2]^2} +
\frac{k_\mu k_\nu (2 k q + q^2)^2}{k^2 [k^2 - M_H^2]^3} -
\right. \nonumber \\[2mm]
& & \left. \frac{k_\mu k_\nu (2 k q + q^2)^3}
{k^2 [k^2 - M_H^2]^3 [(k+q)^2 - M_H^2]} \right]
\end{eqnarray}
Now the last integral is finite at $D=4$ and satisfies the
requirements of the m-theorem. It then goes to zero as $D
\rightarrow 4$ and $\mh \rightarrow \infty$.
The other three terms in $A_{\mu \nu}$
can be evaluated  using standard techniques, and
one gets
\begin{equation}
A_{\mu \nu}  = \frac{i}{16 \pi^2} \left[
g_{\mu \nu} \frac{\mh^2}{4} (\widehat{\Delta}_\epsilon
+ \frac{3}{2}) - g_{\mu \nu} q^2 \frac{1}{12}
(\widehat{\Delta}_\epsilon + \frac{5}{6}) +
q_\mu q_ \nu \frac{1}{3} (\widehat{\Delta}_\epsilon
+ \frac{1}{3}) \right]
\end{equation}

Using the same techniques to evaluate the rest of terms
one finally gets the large $\mh$ contribution of the
$(H-\phi)$ correction to the $W$ self energy
\begin{equation}
\frac{e^2}{s^2} \frac{i}{16 \pi^2} \left[
g_{\mu \nu} \frac{\mh^2}{4} ( \widehat{\Delta}_\epsilon +
\frac{3}{2} ) +
g_{\mu \nu} \frac{\xi M_W^2}{4} (\widehat{\Delta}_\epsilon
+ \frac{3}{2}) - g_{\mu \nu} q^2 \frac{1}{12}
(\widehat{\Delta}_\epsilon + \frac{5}{6}) +
q_\mu q_ \nu \frac{1}{12} (\widehat{\Delta}_\epsilon
+ \frac{4}{3}) \right]
\end{equation}

As a final comment, we would like to point a difference between these
large-m techniques and the commonly used expansion in powers of the
external momenta $q$.  This large-m calculation gives us all the
existing contributions up to an arbitrary power in the external
momenta $q$, as far as they do not vanish in the large-$\mh$ limit.
On the other hand, these techniques provide an extremely easy way to
extract the non-vanishing large mass effects of any loop integral.

\newpage
\setcounter{equation}{0}
\renewcommand{\theequation}{B.\arabic{equation}}
\section*{Appendix B}

In this appendix, we present the results of the various terms
contributing to the matching equations (\ref{MAMA}). Since we are
interested mainly in the differences between the SM terms and
the corresponding ones of the EChL, it is convenient to rewrite
eq.(\ref{MAMA}) in the following form\footnote{ We denote by
$i M $ the direct result from Feynman diagrams}:
\begin{equation}
\left( i M_{abcd}^{{\rm T} \; \mu \nu \rho \lambda} -
i \widehat{M}_{abcd}^{{\rm T} \; \mu \nu \rho \lambda} \right) +
\left( i M_{abcd}^{{\rm C} \; \mu \nu \rho \lambda} -
i \widehat{M}_{abcd}^{{\rm C} \; \mu \nu \rho \lambda} \right) +
\left( i M_{abcd}^{{\rm L} \; \mu \nu \rho \lambda} -
i \widehat{M}_{abcd}^{{\rm L} \; \mu \nu \rho \lambda} \right) = 0
\end{equation}
The tree diagrams contributing to $ ( i M^{\rm T} - i
\widehat{M}^{\rm T}) $ are displayed in fig.(3).  The contributions
from the counterterms are generated  using eq.(\ref{RET}) for the 1PI
Green's functions of the effective theory, $i \widehat{M}^{\rm C}$,
and eq.(\ref{REPS}) for the 1LPI Green's functions of the SM, $i
M^{\rm C}$.  For the difference of counterterms, we use the
following notation:
\begin{displaymath}
\Delta Q \equiv \delta Q - \widehat{\delta Q} \hspace{1.5cm}
{\rm with} \;\; Q = Z_B, Z_W, g^2, {\rm etc...}
\end{displaymath}
The loop contributions to $( i M^{\rm L} - i \widehat{M}^{\rm L} )$
come from the explicit evaluation of all the one loop diagrams
in fig.(3) in the large $\mh$ limit. To perform this calculation
we have used the techniques described in appendix A.

\subsection*{ {\bf $\displaystyle{\gamma \gamma}$}WW }
There are no differences in the tree level contributions
\begin{equation}
\left( i M_{\gamma\gamma WW}^{{\rm T} \; \mu\nu\rho\lambda} -
 i \widehat{M}_{\gamma\gamma WW}^{{\rm T} \;
\mu\nu\rho\lambda} \right) = 0.
\end{equation}
The contributions from the counterterms are
\begin{eqnarray}
\left( i M_{\gamma\gamma WW}^{{\rm C} \; \mu\nu\rho\lambda} -
 i \widehat{M}_{\gamma\gamma WW}^{{\rm C} \; \mu\nu\rho\lambda}
 \right) &=& - i g^4 s^2 \left[ 2 \left(\Delta Z_W - \frac{\Delta
g^2}{g^2} \right)
\frac{1}{g^2}\right]
g^{\mu \nu} g^{\lambda \rho} \nonumber\\
& &+ i g^4 s^2 \left[ \left(\Delta Z_W - \frac{\Delta g^2}{g^2} \right)
\frac{1}{g^2} +  a^b_{11}
\right] \left( g^{\mu \rho} g^{\lambda \nu} + g^{\lambda \mu }
g^{\nu \rho} \right).
\end{eqnarray}
The evaluation of diagrams in fig.(3.a) gives
\begin{eqnarray}
\left( i M_{\gamma\gamma WW}^{{\rm L} \; \mu\nu\rho\lambda} -
 i \widehat{M}_{\gamma\gamma WW}^{{\rm L} \; \mu\nu\rho\lambda}
\right) & = &
\sum_{i=1}^{12} L_i + \sum_{j=1}^3 ( D_j - \widehat{D}_j ) =
\nonumber\\
& & - i \frac{g^4 s^2 }{16 \pi^2} \left[ \frac{1}{6} \left(
 \hat{\Delta}_\epsilon + \frac{5}{6}\right) \right]
 g^{\mu \nu} g^{\lambda \rho} + \nonumber \\[2mm]
& &  i \frac{g^4 s^2}{16 \pi^2} \left[	\frac{1}{12} \left(
 \hat{\Delta}_\epsilon + \frac{4}{3}\right)
\right] \left( g^{\mu \rho} g^{\lambda \nu} + g^{\lambda \mu }
g^{\nu \rho} \right).
\end{eqnarray}

\subsection*{{\bf ${\displaystyle \gamma}$}ZWW}

There are no differences at tree level
\begin{equation}
\left( i M_{\gamma ZWW}^{{\rm T} \; \mu\nu\rho\lambda} -
 i \widehat{M}_{\gamma ZWW}^{{\rm T} \; \mu\nu\rho\lambda} \right) = 0.
\end{equation}
The contributions from the counterterms are
\begin{eqnarray}
\left( i M_{\gamma ZWW}^{{\rm C} \; \mu\nu\rho\lambda} -
 i \widehat{M}_{\gamma ZWW}^{{\rm C} \; \mu\nu\rho\lambda} \right)
& = &
- i g^4 \frac{s}{c} \left[ 2 \left( \Delta Z_W - \frac{\Delta g^2}
{g^2} \right) \frac{c^2}{g^2} + 2 a^b_3 \right]
 g^{\mu \nu} g^{\lambda \rho} \nonumber\\
& &+ i g^4 \frac{s}{c} \left[ \left( (\Delta Z_W - \frac{\Delta
g^2}{g^2})\frac{1}{g^2}
+  a^b_{11}\right) c^2	\right. \nonumber \\
& &\left. \rule[0mm]{0mm}{6mm} + a^b_3 - a^b_{11} + a^b_{12}
\right] \left( g^{\mu \rho} g^{\lambda \nu} + g^{\lambda \mu }
g^{\nu \rho} \right).
\end{eqnarray}
The evaluation of diagrams in fig.(3.b) gives
\begin{eqnarray}
\left( i M_{\gamma ZWW}^{{\rm L} \; \mu\nu\rho\lambda} -
 i \widehat{M}_{\gamma ZWW}^{{\rm L} \; \mu\nu\rho\lambda} \right) &=&
\sum_{i=1}^{18} L_i + \sum_{j=1}^3 ( D_j - \widehat{D}_j ) =  \\
& & - i \frac{g^4 s}{c}\frac{1}{16 \pi^2} \left[ \frac{1}{12}
 \hat{\Delta}_\epsilon ( 2 c^2 + 1 ) + \frac{1}{72}
( 10 c^2 + 17 ) \right]
g^{\mu \nu} g^{\lambda \rho} \nonumber \\
& &+ i \frac{g^4 s}{c}\frac{1}{16 \pi^2} \left[ \frac{1}{12}
 \hat{\Delta}_\epsilon (  c^2 + \frac{1}{2} ) + \frac{1}{36}
( 4 c^2  + \frac{11}{4} ) \right]
 \left( g^{\mu \rho} g^{\lambda \nu} + g^{\lambda \mu }
g^{\nu \rho} \right). \nonumber
\end{eqnarray}

\subsection*{ZZWW}

In this case, there are already differences at tree level since
there is one diagram, $T_1$ in fig.(3c), with a tree level
exchange of a Higgs boson. The large $\mh$ limit of this
diagram gives
\begin{equation}
\left( i M_{ZZWW}^{{\rm T} \; \mu\nu\rho\lambda} -
 i \widehat{M}_{ZZWW}^{{\rm T} \; \mu\nu\rho\lambda} \right) = T_1 =
i \frac{g^2}{c^2} \frac{M_W^2}{\mh^2} g^{\mu \nu} g^{\lambda \rho}.
\end{equation}
The contributions from the counterterms are
\begin{eqnarray}
\left( i M_{ZZWW}^{{\rm C} \; \mu\nu\rho\lambda} -
 i \widehat{M}_{ZZWW}^{{\rm C} \; \mu\nu\rho\lambda} \right) &=&
+ i \frac{g^4}{c^2} \left[ - 2 \left( \Delta Z_W - \frac{\Delta
g^2}{g^2}
\right) \frac{c^4}{g^2} - 4 c^2 a^b_3 - 2 (a^b_5 + a^b_7) + \delta S
\right]
g^{\mu \nu} g^{\lambda \rho} \nonumber\\
& &+ i \frac{g^4}{c^2} \left[ \left( (\Delta Z_W - \frac{\Delta
g^2}{g^2}) \frac{1}{g^2}
+ a^b_{11}\right) c^4 - 2 c^2(-a^b_3+a^b_{11}- a^b_{12} )
\right. \nonumber\\
& & \left. \rule[0mm]{0mm}{6mm} - ( a^b_4 + a^b_6 - a^b_{11}+ 2 a^b_{12})
\right] \left( g^{\mu \rho} g^{\lambda \nu} + g^{\lambda \mu }
g^{\nu \rho} \right).
\end{eqnarray}
where $\delta S$ is given by the following combination of SM counterterms
\begin{displaymath}
\delta S = \frac{M_W^2}{g^2 M_H^2} \left(- \frac{\delta M_H^2}{M_H^2}
+ \frac{ \delta T / v}{M_H^2} + \frac{\delta M_W^2}{M_W^2} \right)
\end{displaymath}
and the SM counterterms $\delta M_H^2, \delta T $ and $ \delta M_H^2$
are given in eq.(\ref{SMCL}).
The evaluation of the diagrams in fig.(3.c) gives
\begin{eqnarray}
\left( i M_{ZZWW}^{{\rm L} \; \mu\nu\rho\lambda} -
 i \widehat{M}_{ZZWW}^{{\rm L} \; \mu\nu\rho\lambda} \right) &=&
\sum_{i=1}^{47} L_i + \sum_{j=1}^6 ( D_j - \widehat{D}_j ) =
\nonumber\\
& & i \frac{g^4 }{c^2}\frac{1}{16 \pi^2} \left[ \frac{1}{6}
 \hat{\Delta}_\epsilon ( - c^4 - c^2 + \frac{43}{4} ) + \frac{1}{36}
( - 5 c^4 - 17 c^2 + 47 ) \right]
g^{\mu \nu} g^{\lambda \rho} \nonumber\\
& &+ i\frac{g^4}{c^2}\frac{1}{16 \pi^2} \left[ \frac{1}{12}
 \hat{\Delta}_\epsilon (c^4 + c^2 - 1)+\frac{1}{72}
(8 c^4+11 c^2 -14)\right] \nonumber \\
& &\left(g^{\mu \rho} g^{\lambda \nu}+g^{\lambda \mu}
g^{\nu \rho}\right).
\end{eqnarray}

\subsection*{WWWW}

The differences at tree level are given by diagrams $T_1$ and $T_2$ of
fig.(3.d). In the large $\mh$ limit we get
\begin{equation}
\left( i M_{WWWW}^{{\rm T} \; \mu\nu\rho\lambda} -
 i \widehat{M}_{WWWW}^{{\rm T} \; \mu\nu\rho\lambda} \right) = T_1 + T_2 =
i g^2 \frac{M_W^2}{\mh^2} \left( g^{\mu \rho} g^{\lambda \nu} +
g^{\lambda \mu } g^{\nu \rho} \right).
\end{equation}
The contributions from the counterterms are
\begin{eqnarray}
\left( i M_{WWWW}^{{\rm C} \; \mu\nu\rho\lambda} -
 i \widehat{M}_{WWWW}^{{\rm C} \; \mu\nu\rho\lambda} \right) &=&
i g^4 \left[ \left( \Delta Z_W - \frac{\Delta g^2}{g^2}
\right) \frac{1}{g^2} + 2 a^b_3 -  a^b_4 - a^b_8 + 2 a^b_9 -
2 a^b_{13} \right]
2 g^{\mu \nu} g^{\lambda \rho} \nonumber\\
& &+ i g^4 \left[ - \left( \Delta Z_W - \frac{\Delta
g^2}{g^2}\right) \frac{1}{g^2}
- 2 a^b_3 - a^b_4 - 2 a^b_5 + a^b_8 - 2 a^b_9
\right. \nonumber\\
& & \left. \rule[0mm]{0mm}{6mm} + 2 a^b_{13} + \delta S
\right] \left( g^{\mu \rho} g^{\lambda \nu} + g^{\lambda \mu }
g^{\nu \rho} \right).
\end{eqnarray}
The one loop diagrams are displayed in fig.(3.d) and their evaluation
in the large $\mh$ limit gives
\begin{eqnarray}
\left( i M_{WWWW}^{{\rm L} \; \mu\nu\rho\lambda} -
 i \widehat{M}_{WWWW}^{{\rm L} \; \mu\nu\rho\lambda} \right) &=&
\sum_{i=1}^{58} L_i + \sum_{j=1}^{12} ( D_j - \widehat{D}_j ) =
\nonumber\\
& &  i \frac{g^4}{16 \pi^2} \left[ \frac{1}{12} \left(
 \hat{\Delta}_\epsilon + \frac{5}{6}\right) \right]
2 g^{\mu \nu} g^{\lambda \rho} \nonumber \\
& &+ i \frac{g^4}{16 \pi^2} \left[  \frac{37}{24}
 \hat{\Delta}_\epsilon + \frac{55}{72}
\right] \left( g^{\mu \rho} g^{\lambda \nu} + g^{\lambda \mu }
g^{\nu \rho} \right).
\end{eqnarray}

\subsection*{ZZZZ}

The differences at tree level are given by diagrams $T_1$, $T_2$
and $T_3$ of fig.(3.e). In the large $\mh$ limit we get
\begin{equation}
\left( i M_{ZZZZ}^{{\rm T} \; \mu\nu\rho\lambda} -
 i \widehat{M}_{ZZZZ}^{{\rm T} \; \mu\nu\rho\lambda} \right) =
T_1 + T_2 + T_3 =
i \frac{g^2}{c^4} \frac{M_W^2}{\mh^2} \left( g^{\mu \nu}
g^{\rho \lambda} + g^{\mu \rho} g^{\lambda \nu} +
g^{\lambda \mu } g^{\nu \rho} \right).
\end{equation}
The contributions from the counterterms are
\begin{equation}
\left( i M_{ZZZZ}^{{\rm C} \; \mu\nu\rho\lambda} -
 i \widehat{M}_{ZZZZ}^{{\rm C} \; \mu\nu\rho\lambda} \right) =
i \frac{g^4}{c^4} \left[ - 2 ( a^b_4 +	a^b_5 ) - 4 ( a^b_6 +
a^b_7  + 2 a^b_{10} ) + \delta S \right]
\left( g^{\mu \nu} g^{\rho \lambda} + g^{\mu \rho}
g^{\lambda \nu} + g^{\lambda \mu } g^{\nu \rho} \right).
\end{equation}

\noindent
The one loop diagrams are displayed in fig.(3.e) and their evaluation
in the large $\mh$ limit gives
\begin{eqnarray}
\left( i M_{ZZZZ}^{{\rm L} \; \mu\nu\rho\lambda} -
 i \widehat{M}_{ZZZZ}^{{\rm L} \; \mu\nu\rho\lambda} \right) &=&
\sum_{i=1}^{75} L_i + \sum_{j=1}^{18} ( D_j - \widehat{D}_j ) =
\nonumber\\
& &  i \frac{g^4}{c^4} \frac{1}{16 \pi^2}  \frac{1}{8} \left[
 13 \hat{\Delta}_\epsilon + \frac{20}{3} \right]
\left( g^{\mu \nu} g^{\lambda \rho} +
 g^{\mu \rho} g^{\lambda \nu} + g^{\lambda \mu }
g^{\nu \rho} \right).
\end{eqnarray}

\newpage

\section*{Figure Captions}
\begin{description}

\item[Fig.1] Feynman rules for the EChL couplings that differ
from the SM. We have shown only those that are relevant for
the present calculation.

\item[Fig.2] One-loop diagrams that give a leading contribution
to the SM combination of counterterms $\delta S$, as explained
in the text.

\item[Fig.3] 1LPI standard model diagrams relevant for the
matching of the four-point Green's functions of gauge fields up
to one loop. We have calculated all the existing diagrams
including gauge, Goldstone boson and Higgs fields in the loops.
We plot here only the diagrams that do not cancel at both sides
of the matching equation, either because they do not exist in
the EChL ($L_i$, $T_i$) or because they are different in the
EChL and the SM ($D_i$).  We have also restricted our plot to
those diagrams that give a non-vanishing contribution to the
matching in the large $\mh$ limit.  Diagrams $D_i$ have to be
calculated also in the effective theory, $\widehat{D}_i$, using
the Feynman rules given in fig.1. \\ All the momenta are taken
incoming and arrows indicate negative charge flux.

{\bf 3.a} Diagrams for the $\gamma \gamma W W$ Green function.
(P1) indicates that the diagram obtained exchanging the
$W^+$ and $W^-$ external legs has also to be included.
There are a total of 12 $L_i$ diagrams and 3 $D_i$ diagrams. \\

{\bf 3.b} $\gamma Z W W$ Green function. (P1) represents the
diagram obtained exchanging the $W^+$ and $W^-$ external legs.
There are 18 $L_i$  and 3 $D_i$ diagrams.\\

{\bf 3.c} $Z Z W W $ Green function.  (P1) represents the
diagram obtained exchanging the $W^+$ and $W^-$ external legs.
There are 47 $L_i$,  6 $D_i$ and 1 $T_i$ diagrams.\\

{\bf 3.d}  $W W W W $ Green function.
(P3) indicates that the three diagrams obtained by exchanging
the ($W^-_\mu \leftrightarrow W^-_\nu $),
($W^+_\rho \leftrightarrow W^+_\lambda $),
and ($W^-_\mu \leftrightarrow W^-_\nu ,
W^+_\rho \leftrightarrow W^+_\lambda $) external legs
have to be also included.
(P3)' indicates the substitutions
($W^+_\rho \leftrightarrow W^+_\lambda $),
($W^+_\rho \leftrightarrow W^-_\nu$) and
($W^-_\mu \rightarrow W^+_\lambda,
W^+_\lambda \rightarrow W^-_\nu,
W^-_\nu \rightarrow W^-_\mu$).
(P1) indicates the exchange
($W^+_\rho \leftrightarrow W^+_\lambda $).
There are 58 $L_i$,  12 $D_i$ and 2 $T_i$ diagrams.\\

{\bf 3.e}  $Z Z Z Z$ Green function.
(P5) indicates the five diagrams obtained by the
following permutations of the $(\mu \nu \rho \lambda)$ external Z's:
$(\mu \nu \lambda \rho), \; (\mu \rho \nu \lambda), \; (\mu \rho
\lambda \nu), \; (\mu \lambda \nu \rho)$ and $(\mu \lambda \rho \nu)$.
(P2) indicates the exchange of $(\mu \nu \rho \lambda)$
by $(\mu \rho \lambda \nu)$ and $(\mu \lambda \nu \rho)$.
There are 75 $L_i$, 18 $D_i$ and 3 $T_i$ diagrams.

\end{description}

\end{document}